\documentclass[12pt]{article}
\usepackage{jheppub}

\newcommand{\be}{\begin{equation}}
\newcommand{\ee}{\end{equation}}
\newcommand{\HH}{\mathcal{H}_H}
\newcommand{\HA}{\mathcal{H}_A}
\newcommand{\HB}{\mathcal{H}_B}
\newcommand{\HR}{\mathcal{H}_R}
\newcommand{\HRB}{\mathcal{H}_{R_B}}
\newcommand{\HRH}{\mathcal{H}_{R_H}}

\newcommand{\lan}{\langle}
\newcommand{\ran}{\rangle}
\newcommand{\Tr}{\mathrm{Tr}}
\newcommand{\textcs}[1]{\textsf{#1}}
\newcommand{\NP}{\textcs{NP}}

\begin{document}
\subheader{Dedicated to John Preskill on the occasion of his 60th birthday}
\title{Quantum Computation vs. Firewalls}
\author[a]{Daniel Harlow}
\author[b]{Patrick Hayden}
\affiliation[a]{Princeton Center for Theoretical Science, Princeton University, Princeton NJ 08540 USA}
\affiliation[b]{School of Computer Science, McGill University}
\emailAdd{dharlow@princeton.edu}
\emailAdd{patrick@cs.mcgill.ca}
\abstract{In this paper we discuss quantum computational restrictions on the types of thought experiments recently used by Almheiri, Marolf, Polchinski, and Sully to argue against the smoothness of black hole horizons.  We argue that the quantum computations required to do these experiments would take a time which is exponential in the entropy of the black hole under study, and we show that for a wide variety of black holes this prevents the experiments from being done.  We interpret our results as motivating a broader type of nonlocality than is usually considered in the context of black hole thought experiments, and claim that once this type of nonlocality is allowed there may be no need for firewalls.  Our results do not threaten the unitarity of of black hole evaporation or the ability of advanced civilizations to test it.  
}
\maketitle
\section{Introduction}
The recently proposed firewall phenomenon \cite{Almheiri:2012rt} has dramatically emphasized the extent to which black holes remain interesting and mysterious in quantum gravity.  Using ``reasonable'' assumptions about the structure of the quantum theory of black holes, the authors of \cite{Almheiri:2012rt}, henceforth referred to as AMPS, have argued that unitarity of black hole evaporation, along with some limited form of locality, is inconsistent with a smooth horizon for an observer falling into an ``old'' black hole.\footnote{We recently found out that a similar argument was made by Braunstein, later joined by Pirandola and \.{Z}yczkowski, back in 2009 \cite{Braunstein:2009my}, who called the firewall an ``energetic curtain''.  Some details of their model do not seem consistent with standard assumptions about black hole physics, however, and restoring those assumptions seems to rule out their curtain-avoiding resolution.}  There has been significant discussion of this claim in the literature, but in our view none of the followup work so far has decisively challenged the original argument.

We will review the AMPS argument in section \ref{firewallsect} below, but a key point in motivating their setup is a claim that an infalling observer is able to extract information from the Hawking radiation of a black hole prior to falling in.\footnote{As emphasized by AMPS, the observer does not actually need to do the experiment to get into trouble.  The possibility of the experiment being done is enough to argue against a smooth horizon.  We will be more precise below about what is meant by ``extract information''.}  The model black hole of Hayden and Preskill \cite{HaydenPreskill}, also partially reviewed below, suggests that this should be possible provided that the observer is able to perform a sophisticated nonlocal measurement on the Hawking radiation that has come out so far.  In this paper we will argue using methods from the theory of quantum computation that this measurement can almost certainly not be done fast enough and thus that the AMPS experiment is not operationally realizable even in principle.     

As a simple example, consider a Schwarzschild black hole in $3+1$ dimensions.  Its entropy is proportional to $M^2$ in Planck units, and it evaporates in a time proportional to $M^3$.  A would-be AMPS experimentalist thus has to extract information from $n\sim M^2$ bits of Hawking radiation in a time $T\sim n^{3/2}$ to be able to jump in before the black hole evaporates.  From a computer science point of view this is very special: the decoding needs to be accomplished in a time that scales as a low-order polynomial in $n$.  In order to get at the information our experimentalist would need to apply a unitary transformation to the Hawking radiation which ``unscrambles'' the desired information by putting it into an easily accessible subfactor of the Hilbert space.\footnote{In line with standard parlance we will sometimes refer to this operation as ``decoding''; we will see the precise connection to what is called decoding in quantum information theory in section \ref{decsect}.}  As we will review below in section \ref{compsec}, applying a generic unitary transformation to an $n$-bit Hilbert space requires time that is exponential in $n$.  Only very special unitary transformations can be implemented faster, and in this paper we will argue that the decoding operation relevant to AMPS is unlikely to be special in this way.  In fact we conjecture, but cannot rigorously prove, that the decoding time for Hawking radiation will in general be exponential in the entropy of the remaining black hole.\footnote{It may seem that beating a small power with an exponential is overkill, but we will show that for more general types of black holes the exponential really is necessary to prevent the AMPS experiment from being done.  In those cases it will be recurrence phenomena rather than evaporation which doom the experiment.} 

In light of our discussion, a firewall enthusiast might nonetheless argue that even though the decoding cannot be done the information is still ``there''.  It has been clear for some time however that operational constraints are important in understanding the structure of the Hilbert space used in describing black hole evaporation, and we view our results in this context.  More concretely the real issue at stake is for which types of situations we should trust effective field theory (EFT).  Traditionally EFT was viewed as holding away from local regions with high energy density or spacetime curvature.  If this were the only way in which EFT could break down however, then we would seem to be led inexorably to information loss \cite{Hawking:1976ra}.  If we wish to maintain belief in unitarity, as AdS/CFT strongly suggests we should, then there must be a more broad set of criteria for when EFT is not be valid.  It is not trivial however to find such criteria which do not flagrantly violate the extraordinary level to which EFT has been experimentally tested.  Careful analysis of thought experiments near black holes in the mid 1990's \cite{sussthorug,sussthor,susspol} led to an additional criterion involving causality: 
\begin{itemize}
\item Two spacelike-separated low-energy observables which cannot both be causally accessed by some single observer do not need to be realized even approximately as distinct and commuting operators on the same Hilbert space.  
\end{itemize}
This criterion was claimed to preclude the apparent contradiction between unitarity and local EFT in Hawking's argument.  It is clear that it does not lead to obvious testable violations of EFT, and it was also claimed to avoid more subtle problems like quantum cloning and unacceptably large baryon number violation.  The key point for us however is that this criterion, which is a profound statement about the structure of the Hilbert space of quantum gravity, was motivated by operational constraints.  It says that whether or not quantum information is ``there'' is indeed related to its practical accessibility.\footnote{The authors of \cite{susspol} also attempted to give a less operational reason for the break down of EFT involving large integrated boosts, although aspects of this claim were later questioned \cite{Polchinski:1995ta}.  It remains to be seen whether a similar argument might exist for the breakdown we advocate here.}    

The deep insight of AMPS is that even with this stronger causal restriction on when we may use effective field theory there is still a paradox that seems to require more modification of the rules, either by having firewalls or by further violating effective field theory.  We interpret our results as supporting a new criterion for the validity of effective field theory: 
\begin{itemize} 
\item Two spacelike-separated low-energy observables which are not both computationally accessible to some single observer do not need to be realized even approximately as distinct and commuting operators on the same Hilbert space.  
\end{itemize}
By computationally inaccessible we mean that one or both of them is so quantum-mechanically nonlocal that measuring it would require more time and/or memory than the observer fundamentally has available.  This criterion clearly implies the previous one, but it is stronger: as we will see, it can apply even if both observables are within the past lightcone of some observer.  In Minkowski space this criterion (and also the causality criterion) is irrelevant, but in spacetimes with singularities there are observers whose available time is fundamentally limited.  This is also true in spacetimes where the fundamental Hilbert space is effectively finite, for example the de Sitter static patch.  In that case recurrence phenomena limit how much time is available for low-energy observation rather than a singularity.\footnote{In \cite{harlowsuss} these limits were used to conjecture that ``precise'' descriptions of spacetime require observers who have access to an infinite amount of information.  This conjecture is distinct from the idea proposed here, but they are clearly related.}  Our proposal violates what AMPS call postulate II, and we will see below that it seems to remove the contradiction that led AMPS to argue for firewalls.  

Lest the reader worry we are throwing the baby out with the bathwater, we here point out that a traditional asymptotic observer at infinity, whom we will refer to as Charlie, has all the time and memory needed to measure the Hawking radiation as carefully as he likes.  Our arguments are thus no threat to the unitarity of black hole evaporation as a precise quantum mechanical process; there will be new restrictions on the validity of effective field theory only for an observer we call Alice who falls into the black hole before it evaporates.\footnote{There could be ``cosmological'' restrictions on what Charlie is able to do, but it seems that these should be decoupled from restrictions ``intrinsic'' to the black hole.  From our discussion of $U_{dyn}$ in section \ref{slowcomp} it seems that in a completely pristine environment Charlie should even be able to test unitarity in polynomial time; we won't address whether or not Charlie would be able to do this in a ``noisy'' environment.}  

Both the weaker ``causality'' criterion and the stronger ``computational'' criterion are negative statements; they tell us what the Hilbert space is \textit{not}.  It is of course very important to understand what the structure of the Hilbert space \textit{is}, and there are two interesting proposals.  The first, sometimes called ``strong complementarity'', takes the point of view that each observer has her own quantum mechanical theory, which is precise for some special observers and approximate in general.\footnote{This general point of view has been advocated by Banks and Fischler, who have tried to realize it more concretely in a formalism called ``holographic spacetime'' \cite{Banks:2001px,Banks:2011av}.  In their setup quantum mechanics is precise for all observers, even those who encounter singularities or recurrences.  They have recently argued that their formalism evades the firewall argument \cite{Banks:2012nn}, but their claim requires a decoupling in Charlie's description of the black hole dynamics between the near-horizon field theory modes and the horizon degrees of freedom. Such a decopuling seems rather implausible, especially in the context of the mining operations of \cite{Brown:2012un,Almheiri:2012rt}.}  There are then consistency conditions between the different theories to ensure that observers who can communicate with each other agree on the results of low-energy experiments visible to them both.  Within this framework it was argued \cite{bousso,harlow} that \textit{if} the AMPS experiment cannot be done, the firewall argument breaks down.\footnote{In \cite{harlow} one of us tried to argue, for a reason having nothing to do with computation, that the experiment cannot be done.  That argument proved unconvincing, and we regard the computational complexity arguments of this paper to be much stronger.}  The basic point is that physical restrictions on what measurements can be done weaken the overlap conditions, allowing for more ``disagreement'' between Alice and Charlie's quantum mechanical descriptions of what is going on.  

The other proposal for the Hilbert space structure, which might be called ``standard complementarity'', claims that there is a single Hilbert space in which states undergo exact unitary evolution.  The quantum physics of various observers are embedded into this single Hilbert space in such a way that each observer has for each time slice a set of distinct operators that approximately commute.  The semiclassical interpretation of an operator according to one observer can be quite different from that of another observer however, and in particular things which are inside the horizon according to one observer might be outside from the point of view of another.  This viewpoint is essentially that of \cite{sussthorug,sussthor,susspol}, and in the context of firewalls it is sometimes called ``$A=R_B$'' for reasons we will soon see.  It has been suggested by several people\footnote{Including but probably not limited to \cite{bousso, Papadodimas:2012aq, Jacobson:2012gh}.} as a way out of firewalls, but it has so far run into various paradoxes involving apparent cloning and acausality \cite{bousso,Susskind:2012uw}.  We postpone further discussion of these two options until after we present the firewall argument, but it seems that in either framework our computational breakdown of EFT may be sufficient to avoid the paradoxes without any need for firewalls.\footnote{Another interesting proposal, raised again recently in the firewall discussion by John Preskill and Alexei Kitaev, is the Maldacena/Horowitz ``black hole final state'' scenario \cite{Horowitz:2003he}.  Our results could have interesting implications for that idea, but we won't explore them here.}  

It is also interesting to think about whether more general types of black holes have firewalls.  For example Reissner-Nordstrom black holes semiclassically seem to take an infinite amount of time to evaporate, apparently allowing ample time for quantum computation prior to jumping in.  We will explain however that the well-known ``fragmentation'' phenomenon \cite{Preskill:1991tb,maldstrom} destroys the black hole well before the computation can be completed.  Big AdS black holes do not evaporate at all, so the AMPS argument does not directly apply to them, but arguments have been put forward suggesting that they nonetheless have firewalls.  In particular Don Marolf has argued that one could simply mine the black hole until half of its entropy is gone, after which the mining equipment would play the role of the Hawking radiation in the original AMPS argument.  If the decoding time is indeed exponential in the entropy of the black hole, as we argue it is, then it becomes comparable to the Poincar\'e recurrence time of the AdS-Schwarzschild space \cite{sussrec}; we argue that no observer or computer can isolate itself from a big black hole for so long in AdS space.  

The strictest test of our criticism of the AMPS experiment uses a setup suggested to us by Juan Maldacena, in which a large AdS black hole is placed very far down a throat whose geometry is asymptotically Minkowski. By putting the decoding apparatus out in the Minkowski region, it seems that one could use the redshift to arbitrarily speed up the decoding time compared to the evaporation time.  We will explain in section \ref{adssect} however that the decoupling which makes the decay slow in this situation also makes it very difficult to send the results of the computation back down the throat, and for a particular example we show that, for a wide variety of probes, the time required to successfully send a message down the throat is longer than the recurrence time of the black hole down the throat.  So indeed it seems there is a fairly robust conspiracy preventing the AMPS experiment from being done.  These results are consistent with a point of view expressed by Aaronson \cite{Aaronson:2005qu} that the laws of physics should not permit computational machines that radically alter the basic structure of complexity theory. At most, they should force some marginal changes around the edges, as in the case of Shor's factoring algorithm.

It is interesting to note that if our computational complexity argument is correct, it supersedes many of the classic black hole thought experiments \cite{sussthorug,sussthor,susspol,HaydenPreskill}.  In particular the argument of \cite{HaydenPreskill} that the scrambling of information by a black hole in a time no faster than $M\log M$ is necessary to prevent observable cloning would be no longer be needed.    

An objection to our argument that we have sometimes encountered is the following: say that we have a quantum computer which is simulating some quantum gravity theory like $\mathcal{N}=4$ Super Yang-Mills theory.  Somebody who is working the computer could decide to make a black hole in the simulation, let it evaporate for a while, pause the simulation, use the computer to decode the radiation while the simulation is paused, and then start the simulation again.  In this way divine intervention from the outside is able to circumvent our claim that the radiation cannot be decoded in time for an infalling observer to see it.  One can try to make this proposal sound somewhat less artificial by setting it in AdS, where the claim would be that, since the spacetime is not globally hyperbolic, these manipulations can be done by coupling the bulk to an arbitrary external system at the boundary and manipulating this coupling at will.  From our point of view however the main question at stake here is whether or not nonsingular initial data on some Cauchy surface generate singularities (such as firewalls) beyond those predicted by low energy physics.  We are less concerned about the question of what happens if some godlike entity, external to the system, disrupts it in some arbitrary way.  In particular if something subtle like standard complementarity is going on in the structure of the Hilbert space, then outside manipulations which naively don't affect the region behind the horizon might actually have rather drastic effects on it.\footnote{Adam Brown has pointed out that somebody who believes this ``pause the simulation argument'' might also use it to support Hawking's original nice-slice argument for non-unitarity by claiming that an external deity could reach inside of the black hole and teleport the collapsing matter that originally made the black hole out into the vicinity of the Hawking radiation, producing verifiable cloning unless unitarity is false.  This clearly illustrates the danger of using a naive picture of the Hilbert space in ``pause the simulation'' arguments.  Another example is the eternal two-sided AdS black hole \cite{Maldacena:2001kr}, where even though the dual is two decoupled CFTs it seems reasonable to expect that doing arbitrary manipulations on one of the CFTs destroys the interior for somebody who jumps in from the other side.}  We would be more concerned by this argument if the coupling of AdS to the external system were accomplished in a manner in which it were clear that low energy physics might be expected to hold throughout the joint system, and indeed one can interpret the throat geometry we discuss in section \ref{adssect} as a realization of this.  In that setup physical restrictions seem to prevent any observer from overcoming the computational complexity of decoding.  

This introduction has telegraphically sketched our main points. In the remainder of the paper we will make the case again in much more depth.  Because of our expected audience we will try to keep our discussions of quantum computation and coding self-contained, but the same definitely cannot be said for our discussions of black holes and gravity.  Other work on firewalls includes \cite{Nomura:2012sw,Mathur:2012jk,Bena:2012zi,Giveon:2012kp,Chowdhury:2012vd,Avery:2012tf,Hossenfelder:2012mr,Hwang:2012nn,Larjo:2012jt,Page:2012zc,Giddings:2012gc}.

\section{The Firewall Argument}\label{firewallsect}
We begin with a somewhat reorganized presentation of the original argument of \cite{Almheiri:2012rt}.  Their argument rests on some basic assumptions about the quantum description of a black hole from the points of view of an external observer Charlie and an infalling observer Alice, and we will try to be clear about what these assumptions are.  The argument has many fine technical points, and we will not address all of them.  Our goal is to motivate equation \eqref{Psi}, on which the rest of our paper will be based.  

\subsection{A Quantum Black Hole from the Outside}
We'll first discuss Charlie's description, which is based on the following three postulates:
\begin{itemize}
\item According to Charlie, the formation and evaporation of the black hole is a unitary process.  Moreover, in addition to an asymptotic S-matrix, we can also think about either continuous or discrete time evolution in which at any given time there is a pure quantum state $|\Psi\ran$ in some Hilbert space $\mathcal{H}_{outside}$.
\item At any given time in this unitary evolution we can factorize $\mathcal{H}_{outside}$ into subfactors with simple semiclassical interpretations:
\be\label{outfactor}
\mathcal{H}_{outside}=\HH \otimes \HB \otimes \HR.
\ee
Here $\HR$ are the modes of the radiation field outside of the the black hole, roughly with Schwarzschild coordinate radius $r>3GM$.  $\HB$ are the field theory modes in the near-horizon region, roughly with support over $2GM+\epsilon<r<3GM$ where $\epsilon$ is some UV cutoff.  The geometry in this region is close to Rindler space.  $\HH$ are the remaining degrees of freedom in the black hole, which we can heuristically think of as being at the stretched horizon at $r=2GM+\epsilon$.  Clearly the distinctions between these subfactors are somewhat arbitrary. In particular, it will be convenient to restrict the modes in $\HB$ to have Schwarzschild energy less than the black hole temperature $T=\frac{1}{4\pi G M}$.  Those with higher energy are not really confined to the near-horizon region and we will include them as part of $\HR$. The time evolution of $|\Psi\ran$ does not respect this factorization and cannot be computed using low energy field theory, but for our purposes it is enough to consider the state at a given time.
\item If $|H|$ and $|B|$ are the dimensionalities of $\HH$ and $\HB$ respectively, then $\log|H|$ and $\log|B|$ are both proportional to the area of the black hole horizon in Planck units at the time at which we study $|\Psi\ran$.  Thus their size decreases with time.  Naively $\HR$ is infinite dimensional since all sorts of things could be going on far from the black hole, but we will restrict its definition to only run over the subfactor which in $|\Psi\ran$ is nontrivially involved in the black hole dynamics.  Thus the size of $\HR$ grows with time.
\end{itemize}
The third assumption leads to an interesting distinction between ``young'' and ``old'' black holes \cite{HaydenPreskill}, with the separation based on whether $|R|$ is bigger or smaller than $|H||B|$.   When the black hole is young, $|R|$ is quite small and $B$ and $H$ are entangled significantly.  As the black hole becomes old however, $|R|$ becomes large and $B$ and $H$ taken together become a small subsystem of the full Hilbert space $\mathcal{H}_{outside}$.  Page's theorem \cite{pagethm} then suggests that the combined system $BH$ has a density operator which is close to being proportional to the identity operator:\footnote{Page's theorem says that in a Hilbert space which can be factorized into $\mathcal{H}_A\otimes \mathcal{H}_B$, with $|A|\leq|B|$, a typical pure state has $S_A=\log|A|-\frac{|A|}{2|B|}+O\left(\frac{1}{|A|}\right)$.}
%
\be\label{rhoBH}
\rho_{BH}\approx \frac{1}{|B||H|}I_B\otimes I_H.
\ee  
The time beyond which this is true has come to be called the Page time.  More carefully we would expect a thermal distribution in the Schwarzschild energy at the usual temperature $T=\frac{1}{4\pi G M}$, but since we have put high-frequency modes in $\HR$ the thermal density matrix for $\HH\otimes\HB$ is quite close to \eqref{rhoBH}.\footnote{
It is important in what follows that these ``low-energy'' modes can have quite high proper energy near the horizon, so they are relevant to the experience of an observer who is near the horizon even though the Schwarzschild temperature is typically very small compared to any scale relevant to that observer.}  

We can describe the state concisely by saying that $BH$ is maximally entangled with a subspace in $R$.  More precisely, there is a $|\Psi\ran$-dependent decomposition of $\HR$
\be
\HR=\left(\HRH\otimes\HRB\right)\oplus \mathcal{H}_{other},
\ee
with $|R_H|=|H|$ and $|R_B|=|B|$, such that we can write the state of the full system, to a good approximation, as\footnote{This representation of the state is called the Schmidt decomposition.}
\be\label{Psi}
|\Psi\ran=\left(\frac{1}{\sqrt{|H|}}\sum_h|h\ran_H |h\ran_{R_H}\right)\otimes \left(\frac{1}{\sqrt{|B|}}\sum_b|b\ran_B |b\ran_{R_B}\right).
\ee
Here $h$ and $b$ label orthonormal bases for $\HH$ and $\HB$ respectively, and we have chosen convenient complementary bases for $\HRH$ and $\HRB$.  $R_H$ and $R_B$ are called the \textit{purifications} of $H$ and $B$ respectively.  The state has zero projection onto $\mathcal{H}_{other}$.  

This form of $|\Psi\ran$ makes it clear that any measurement done on $B$ is perfectly correlated with some other measurement done on $R_B$.  This consequence of the entanglement was emphasized in \cite{Almheiri:2012rt}; in their language measurements done on $R_B$ project onto particular states in the basis $|b\ran_B$.  We hope that it is clear however that the presence of this entanglement does not \textit{require} any such measurement to be done; once we accept the three assumptions the entanglement follows directly.  Indeed we would argue that Charlie's ability to measure $R_B$ provides justification for accepting the Hilbert space structure of the model.\footnote{Remember Charlie stays outside the black hole and has an arbitrarily large amount of time and resources, so there seems to be no limit on his experimental ability.  This will be different for Alice, whom we discuss now.}

\subsection{A Quantum Black Hole from the Inside}
\begin{figure}
\begin{center}
\includegraphics[height=8cm]{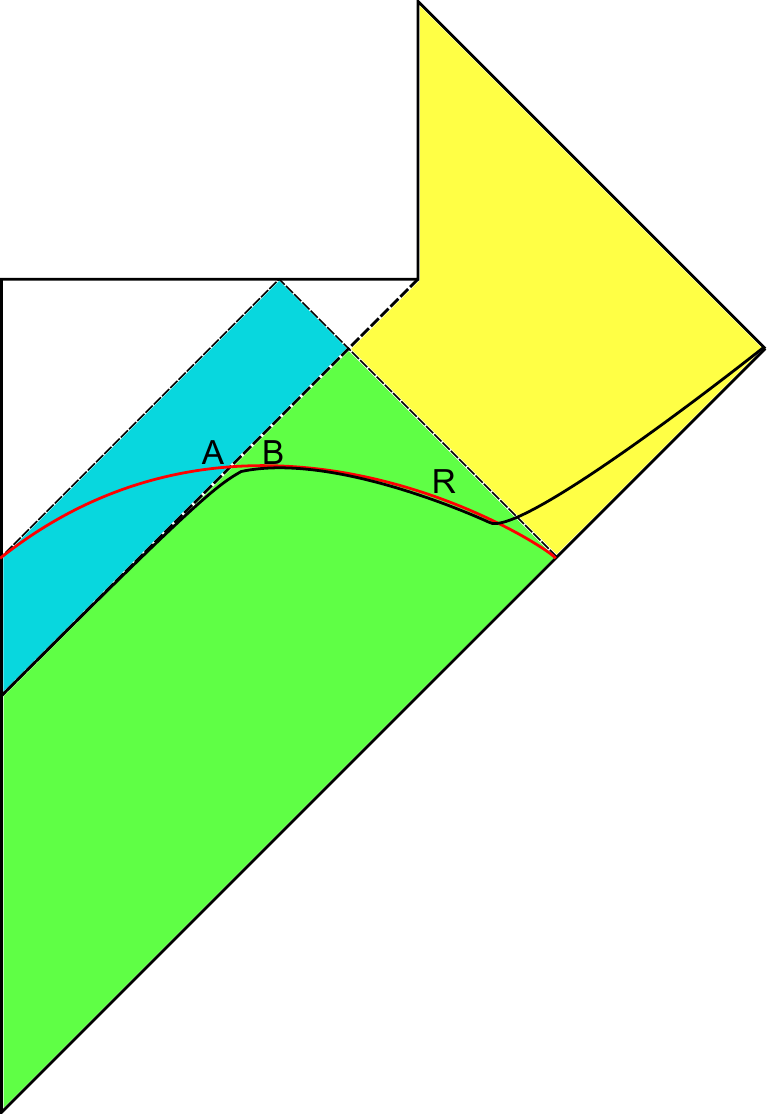}
\end{center}
\caption{Alice's quantum mechanics, compared to Charlie's.  The world inside her horizon is drawn in blue and the time slice she quantizes on is in red.  For reference Charlie's world is in yellow, and the overlap is green.  We've chosen Charlie's black slice to coincide closely with Alice's near $B$ and $R$.}\label{alice}
\end{figure}

We now consider Alice the infalling observer's point of view.  As mentioned in the introduction, it may be possible to ``embed'' Alice's quantum mechanics into Charlie's via some sort of nontrivial operator mapping.  We will discuss this eventually but for the moment will just treat Alice's theory as an independent construction.  Here are the basic assumptions about it:
\begin{itemize}
\item Although Alice eventually hits the singularity, we imagine that well before that she has an approximately quantum mechanical description of her experiences in terms of a quantum state on a time slice like the one shown in figure \ref{alice}.  We will not insist that the state be pure.

\item Alice's Hilbert space also has a roughly semiclassical factorization of the form
\be
\mathcal{H}_{inside}=\HA\otimes\HB\otimes\HR \otimes\mathcal{H}_{H'}.
\ee
Here $\HB$ and $\HR$ are factors shared with Charlie, since they are outside the black hole horizon and are causally accessible to both Charlie and Alice.  $\HA$ are the field theory modes just inside the horizon, say with support over $GM<r<2GM-\epsilon$.  $\mathcal{H}_{H'}$ are the remaining degrees of freedom having to do with Alice's horizon (which is distinct from the black hole horizon).  $\HH$ is absent; the region $2GM-\epsilon<r<2GM+\epsilon$ is passed through by Alice in an extremely short period of time and does not have any operational meaning to her.  Of course, at times long before she falls in, the black hole horizon is indistinguishable from her horizon, and $\HH$ is roughly part of $\mathcal{H}_{H'}$.  We emphasize however that the details of this accounting don't matter.

\item Because $\HB$ and $\HR$ are shared with Charlie, we must have $\rho^{\{Alice\}}_{BR}=\rho^{\{Charlie\}}_{BR}$.  This is sometimes called the overlap rule, and it is designed to prevent contradictions where Alice and Charlie disagree about the results of experiments they can communicate about.
\end{itemize}

With these assumptions, one can now argue following AMPS that Alice must not see a smooth vacuum at the horizon.  Recall that the modes in $\HB$ and $\HA$ are basically Rindler modes on two different sides of a Rindler horizon.  In the Minkowski vacuum state such modes are close to maximally entangled:
\be\label{smoothvac}
|vac\ran\sim\sum_\omega e^{-\frac{\beta\omega}{2}}|\omega\ran_A|\omega\ran_B,
\ee
where $\beta\omega$ is the dimensionless Rindler energy; here it is just the ratio of Schwarzschild energy to the black hole temperature.  This however is problematic from our discussion of Charlie.  We argued that for an old black hole Charlie should see $B$ being close to maximally entangled with $R_B$, and by the third assumption about Alice this must also be true for her.  But entanglement in quantum mechanics is monogamous: such entanglement prevents $B$ from also being entangled with $A$ as in \eqref{smoothvac}.  One way to see this more precisely, again following AMPS, who themselves were motivated by a similar argument due to Mathur \cite{Mathur:2009hf}, is to note that strong subadditivity~\cite{lieb1973proof} requires
\be\label{strongsub}
S_{ABR_B}+S_B\leq S_{AB}+S_{BR_B}.
\ee
By construction we have $S_{BR_B}=0$, which also implies (for example by the Araki-Lieb triangle inequality) that $S_{ABR_B}=S_A$.  For \eqref{strongsub} to be consistent with the subadditivity inequality $S_A+S_B\geq S_{AB}$ we then must have
\be
I_{AB}\equiv S_A+S_B-S_{AB}=0.
\ee 
$I_{AB}$ is called the \textit{mutual information} between $A$ and $B$, and it is zero if and only if the density matrix $\rho_{AB}$ is actually a product $\rho_A\otimes \rho_B$.  This is clearly inconsistent with $A$ and $B$ being close to the highly entangled state \eqref{smoothvac}.  This concludes the AMPS argument; one possible interpretation is that the resolution of the contradiction is that there is a ``firewall'' of high energy quanta at the horizon of an old black hole which annihilates any infalling observer.

\subsection{A Way Out?}
A key step in the AMPS argument is that $B$ and $R_B$ are accessible to both Charlie and Alice and that therefore they must agree on the entanglement between them.  But is this really true?  In \cite{harlow} it was argued that it is difficult for Alice to measure $B$ because she passes through it quickly, but the details of that argument have not worked out satisfactorily. The much more difficult measurement however is the one on $R_B$.  $R_B$ is defined as the subfactor of the Hawking radiation which is entangled with $B$, but this subfactor is presumably very convoluted from the point of view of a basis of Hawking quanta that is easy to measure.  Probing $R_B$ entails doing quantum measurements involving nonlocal quantum superpositions of large numbers of Hawking quanta.  Such measurements need to be very carefully engineered, and one might expect that this engineering takes a significant amount of time.  This is no problem for Charlie, who has all the time in the world to look at the Hawking radiation, but Alice needs to be able to make the measurement fast enough that she can then jump into the black hole before it evaporates.  In this paper we will argue that Alice simply does not have time to do this.  

As a result one can consider modifying her postulated quantum mechanics, for example in one of the two directions described in the introduction.  In the ``strong complementarity'' approach, Alice's quantum mechanics has no direct relation to Charlie's.  They are related only insofar as they must agree on the results of experiments which are visible to both of them.  Since operators acting on $\HRB$ are not accessible to Alice, our computational criterion for the breakdown of effective field theory allows us to either disentangle $\HRB$ from $\HB$ in Alice's theory (but not Charlie's) or more perhaps simply to just remove $\HRB$ from her Hilbert space altogether.  This then ``frees up'' $B$ to be entangled with $A$, ensuring Alice a smooth journey across the horizon.  
In ``standard complementarity'' we instead think of Alice's theory as being embedded in Charlie's.  To avoid firewalls one might try to arrange that operators on Alice's $\HA$ are really just her interpretations of operators acting on what Charlie would have called $\HRB$.  As mentioned in the introduction, this idea, which is essentially an enhanced version of the original proposal of \cite{sussthorug,sussthor,susspol}, is referred to as ``$A=R_B$'' since the entanglement will only work out consistently if we ``build'' interior operators out of exterior operators which already have the correct entanglement with operators acting on $\HB$.  Without limitations on Alice's ability to directly measure $R_B$ however it seems to lead to paradoxes and has thus been viewed with some skepticism.  We view our work as potentially restoring the credibility of this proposal.  We will discuss this a bit more concretely in section \ref{options} below.

\section{The AMPS Experiment as a Quantum Computation}\label{sec3}

We now begin our discussion of the decoding problem confronting Alice.  We will phrase the discussion in terms of Charlie's Hilbert space, since for the moment we are following AMPS and granting that Charlie and Alice must agree on the density matrix of $\HB\otimes \HR$.  For simplicity we will throughout model all Hilbert spaces using finite numbers of qubits.  In the Schmidt basis the state of the old black hole she is interested in is given by equation \eqref{Psi}, but this basis is very inconvenient for discussing Alice's actions.  From here on we will use exclusively a basis for the radiation field which is simple for Alice to work with, and whose elements we will write as
\be\label{Rbasis}
|bhr\ran_R\equiv |b_1\ldots b_k,h_1\ldots h_{m}r_1\ldots r_{n-k-m}\ran_R.
\ee
Here there are $n\equiv \log_2 |R|$ total qubits, each of which we assume Alice can manipulate easily.  $b_1\ldots b_k$ are the first $k$ of these qubits, where $k$ is the number of bits in $\HB$, and $m$ is the number of bits in $\HH$.  We can think of $k+m$ as the number of qubits remaining in the black hole; we will eventually argue that the decoding time is likely of order $2^{k+m+n}$.  The $r_i$ qubits make up the remainder of the modes which have non-trivial occupation from the Hawking radiation.  Roughly we might expect that
\be\label{entguess}
n\approx S_{initial}/\log 2-k-m,
\ee
where $S_{initial}$ is the horizon area in Planck units of the original black hole prior to any evaporation.  This is something of an underestimate because the computational basis is local while the information is nonlocal, but this ``coarse-graining'' enhancement isn't large.  The radiation mostly comes out in $s$-wave quanta so it is effectively one-dimensional.  It extends out to a distance $L\sim M^3$ and consists mostly of quanta whose energy is of order $1/M$, so its thermal entropy is $n\approx L T\sim M^2$, which is still of order the initial black hole entropy as we would conclude from \eqref{entguess}.\footnote{We thank Don Page for several useful discussions of this point.}  Perhaps surprisingly the black hole makes quite efficient use of the information storage capacity available to it.  

We will adopt standard terminology and refer to the basis \eqref{Rbasis} as the computational basis.  In the computational basis we can write the state \eqref{Psi} as
\be\label{compPsi}
|\Psi\ran=\frac{1}{\sqrt{|B||H|}}\sum_{b,h}|b\ran_B |h\ran_H U_R|bh0\ran_R,
\ee
where $U_R$ is some complicated unitary transformation on $\HR$.  What unitary transformation it is will depend on the details of quantum gravity, as well as the initial state of the black hole.  For simplicity we have defined it to act on the state where all of the $r_i$ qubits are zero.  Clearly the challenge Alice faces is to apply $U_R^{\dagger}$ to the Hawking radiation, after which it will be easy for her to confirm the entanglement between $\HB$ and $\HRB$.  Engineering a particular unitary transformation to act on some set of qubits is precisely the challenge of quantum computation, and we will henceforth often refer to Alice's task as a computation.

So far we have been interpreting $\HB$ as the thermal atmosphere of the black hole, but to actually test the AMPS entanglement it would be silly for Alice to try to decode all of the atmosphere.  Indeed the separation between $\HB$ and $\HH$ is rather ambiguous, and we are free to push some of the atmosphere modes we are not interested in into $\HH$.  So from here on we will mostly take $k$ to be $O(n^0)$.  This ostensibly simplifies her computation, because in any event she only needs to implement $U_R$ up to an arbitrary element of $U(2^{n-k})$ acting on the last $n-k$ qubits of the radiation.  This simplification turns out to be irrelevant since the dimensionality of the coset $U(2^n)/U(2^{n-k})$ still scales like $2^{2n}$ for $k>0$, but more importantly it is clear from \eqref{compPsi} that $U_R$ has only really been defined acting on the $2^{k+m}$-dimensional subspace of $\HR$ spanned by states of the form $|bh0\ran$.  Its action on the orthogonal subspace can be chosen freely to simplify the computation provided that it preserves orthogonality, and we will see below that this reduces the time of Alice's computation from a naive $2^{2n}$ to only $2^{k+m+n}$.  Since the black hole is old we have $n>k+m$, so this is at least $2^{2(k+m)}$.

Since the unitary group is continuous it is clear that Alice will not be able to do the computation exactly.  We thus need a good definition of how ``close'' she needs to get to reliably test the entanglement.  One standard way to quantify closeness of operators is the trace norm \cite{Fuchs:1995mk}, which for an operator $A$ is defined as
\be
||A||_1\equiv \Tr\left(\sqrt{A^\dagger A}\right).
\ee
When $A$ is hermitian this is just the sum of the absolute values of its eigenvalues.  The motivation for this definition is as follows: say $\rho_1$ and $\rho_2$ are two density matrices, and $\Pi_a$ is a projection operator for some measurement to give result $a$.  Then
\be
|P_1(a)-P_2(a)|=|\Tr\left\{(\rho_1-\rho_2)\Pi_a\right\}|\leq||(\rho_1-\rho_2)||_1,
\ee
so if the trace norm of the difference of two density operators is less than $\epsilon$ then the probabilities they predict for any experimental result will differ by at most $\epsilon$.  The trace norm of their difference is clearly preserved by unitary evolution.  

If both states are pure then the trace norm of their difference has a simple interpretation.  For any two pure states $|\Psi_1\ran$ and $\Psi_2$ we can write that
\be
|\Psi_2\ran=e^{i\alpha}\left(\sqrt{1-\delta^2/4}|\Psi_1\ran+\frac{\delta}{2} |\chi\ran\right),
\ee
where $|\chi\ran$ is orthogonal to $|\Psi_1\ran$, $\alpha$ is real, and $\delta$ is real and positive.  A simple calculation then shows that
\be\label{puretrnorm}
||\,|\Psi_2\ran\lan\Psi_2|-|\Psi_1\ran\lan\Psi_1|\,||_1=\delta.
\ee


\subsection{Quantum Computing is Hard}\label{bigcompsect}
\begin{figure}
\begin{center}
\includegraphics[height=6cm]{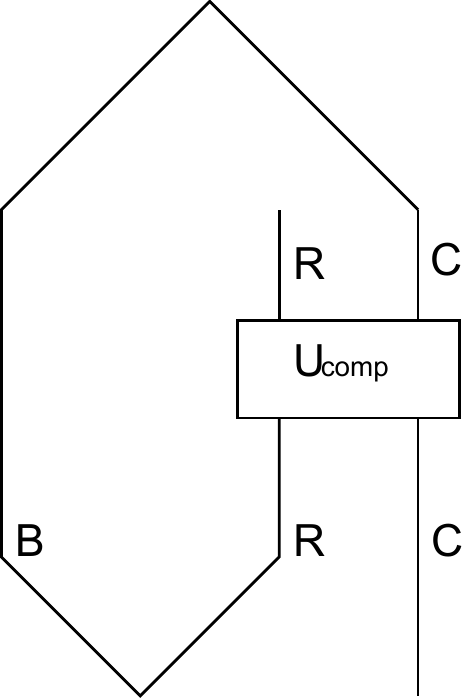}
\end{center}
\caption{What the computer does.  The connecting lines at the top and bottom indicate entanglement, and time goes up.  The subsystem $H$ goes along for the ride, and after the computation its purification is split between $R$ and $C$ in some complicated way.}\label{bigcomp}
\end{figure}

We begin with a rather formal discussion of Alice's computation to illustrate some basic limitations on what is possible; we will take a more standard approach in the following subsection.
In order to do her computation Alice needs to adjoin the radiation to some computer, whose initial state lives in a new Hilbert space $\mathcal{H}_C$, and then wait for the natural unitary evolution $U_{comp}$ on $\HR\otimes\mathcal{H}_C$ to undo $U_R$ and put the bits which are entangled with $B$ into an easily accessible form, let's say the first $k$ qubits of the memory of the computer.  We show this pictorially in figure \ref{bigcomp}.
For most of this subsection we will fix the amount of time that the computer runs for, meaning that we will take $U_{comp}$ to be determined by the laws of physics and thus unchangeable.  The only way Alice has any hope of getting the computer to do what she wants is by carefully choosing its initial state.  Without loss of generality she can take this initial state to be pure, perhaps at the cost of increasing the size of the computer.  We will show that no matter how large she makes her computer, it is very unlikely that she will be able to find even one initial state which does the computation.  

More precisely what Alice would like to do is find a state $|\Psi\ran_{C}$ which for all $b$ and $h$ evolves as
\be \label{comptask}
U_{comp}:U_R|bh0\ran_R\otimes |\Psi\ran_{C}\mapsto|\mathrm{something}\ran\otimes|b\ran_{mem},
\ee
where $|\mathrm{something}\ran$ is any pure state of the computer and radiation minus the first $k$ bits of the memory.  $|\mathrm{something}\ran$ can and will be different for different $h$.  To estimate how likely it is that such a $|\Psi\ran_{C}$ exists, we discretize the Hilbert space using the trace norm.  In any Hilbert space $\mathcal{H}$ of dimension $d$ we can find a finite set $S_\epsilon \subset \mathcal{H}$ with the property that any pure state in $\mathcal{H}$ is within trace norm distance $\epsilon$ of at least one element of $S_\epsilon$.  Such a set is called an $\epsilon$-net, and it is not too hard to get an estimate of how many elements it must have \cite{hayden2004randomizing}.  One first observes from \eqref{puretrnorm} that half of the trace norm difference is weakly bounded by the Hilbert space norm:
\be\label{norms}
|| |\Psi_2\ran-|\Psi_1\ran ||_2^2=2\left(1-\cos \alpha \sqrt{1-\delta^2/4}\right)\geq \left(\frac{\delta}{2}\right)^2=\left(\frac{1}{2}||\,|\Psi_2\ran\lan\Psi_2|-|\Psi_1\ran\lan\Psi_1|\,||_1\right)^2.
\ee
Thus an $\epsilon/2$-net for the Hilbert space norm is also an $\epsilon$-net for the trace norm.  The minimal size of an $\epsilon/2$-net for the Hilbert space norm is the number of balls of radius $\epsilon/2$ centered on points on the unit sphere in $\mathbb{R}^{2d}$ that are needed to cover it, which at large $d$ is proportional to some small power of $d$ times $\left(\frac{2}{\epsilon}\right)^{2d-1}$.\footnote{This is a slight overestimate for the size of the trace norm $\epsilon$-net because the Hilbert space norm distinguishes between states that differ only by a phase while the trace norm does not.  We can fix this by taking the quotient of the unit $\mathbb{S}^{2d}$ by the phase to get to a unit $\mathbb{CP}^{d-1}$, whose volume is just a negligible power in $d$ times the volume of $\mathbb{S}^{2d}$.  This quotient effectively sets $\alpha=0$ between any two states, in which case equation \eqref{norms} tells us that the Hilbert space norm becomes close to one half the trace norm.  The induced metric $\mathbb{CP}^{d-1}$ inherits from $\mathbb{R}^{2d}$ will then for small $\epsilon$ be the same as one half the trace norm distance.  The upshot is then that the number of balls needed to cover $\mathbb{CP}^{d-1}$ scales like $\left(\frac{2}{\epsilon}\right)^{2d-2}$.}  Intuitively we may just think of unitary evolution as an inner-product preserving permutation of the $\left(\frac{2}{\epsilon}\right)^{2d}$ states.  

Applying this now to our discussion of the computer, for fixed $b$ and $h$ the total number of possible states that could appear on the right hand side of \eqref{comptask} is $\left(\frac{2}{\epsilon}\right)^{2|C||R|}$.  The number of possible $|\mathrm{something}\ran$'s is $\left(\frac{2}{\epsilon}\right)^{2|R||C|2^{-k}}$.  For a given $|\Psi\ran_C$, the probability over random choices of $U_{comp}$ that \eqref{comptask} holds for all $2^{k+m}$ values of $b$ and $h$ is then roughly $
\left(\frac{2}{\epsilon}\right)^{-2|C||R|2^m(2^k-1)}$.  The number of available initial states is $\left(\frac{2}{\epsilon}\right)^{2|C|}$, so the probability that Alice can find one for which \eqref{comptask} holds is no more than\footnote{In this counting we can easily ignore the constraint that orthogonal states must be sent to orthogonal states.}
\be\label{compprob}
P=\left(\frac{2}{\epsilon}\right)^{-2|C|\left(|R|2^m(2^k-1)-1\right)}.
\ee
For any nontrivial $k$ and $|R|=2^n$, it is clear that this probability is extraordinarily small.  Making the computer bigger just makes it even more unlikely that the computation can be done!  

What then is Alice to do?  One might hope that, although the probability of success is small for any given computer size, by searching over many values of $|C|$ Alice might find one that works.  This is a bad idea; summing \eqref{compprob} over $|C|$ produces a finite sum whose value remains exponentially small in $|R|$.  Alice can do better however by varying the running time of the computer.  This lets her sample a variety of $U_{comp}$'s without increasing the size of the Hilbert space.  If each $U_{comp}$ is different, then the longest she might have to wait to get a $U_{comp}$ that works is 
\be\label{quantrec}
t\sim e^{2\log\left(\frac{2}{\epsilon}\right)|R||C||H||B|}.
\ee
Although finite, this is unimaginably long for any reasonable system size.  For an astrophysical black hole in our universe it is something like $10^{10^{10^{40}}}$ years.  

The timescale \eqref{quantrec} has a simple physical interpretation; it is the quantum recurrence time.  This is the timescale over which a quantum system comes close to any given quantum state, and as we found here is double-exponential in the entropy of the whole system.  In doing the computation this way, Alice is simply waiting around for a quantum recurrence to do it by pure chance.

Fortunately for civilization, these simple estimates are not the final word on quantum computing power.  In particular, \eqref{quantrec} does not really hold unless $U_{comp}$ is chosen randomly at each time step.  To the extent that there is some structure in how $U_{comp}$ varies with time, as there is in our world, Alice can take advantage of it to speed up her computation.  Similarly, if the way $U_{comp}$ changes with increasing $|C|$ also has structure, she can use that as well.  The lesson of this section however is that without using special properties of the computer-radiation dynamics, no amount of preparation of the initial state of her computer will allow Alice to do her computation in any reasonable amount of time. In the following two sections we will see that by using such physical properties Alice is able beat the double exponential in computer entropy down to a single exponential in just the radiation entropy, but we will also argue that that is probably all she gets.

As a tangential comment it is interesting to note that the result of this section is actually special to quantum mechanics; there is a somewhat analogous problem in classical coding which can easily be solved by making the computer bigger.  Say that we have a classical bit string of length $n$.  There are $2^n$ such strings, but say we are interested in some subset of size $2^k$.  For example this could be the set with a $k$-bit message in the first $k$ bits and zero for the rest.  Acting with some random permutation on the space of $2^n$ strings, we can send these $2^k$ strings to a set of $2^k$ scrambled ``code words'', which are analogous to some basis for $R_B$ in the quantum problem.  We could then imagine adjoining one of our $n$-bit strings to a $c$-bit ``computer'' string, and then acting with a given permutation of the $2^{n+c}$ states of this larger system.  This permutation is the analogue of our $U_{comp}$.  The question is then the following: given this larger permutation, can we find a single initial string for the computer such that, after the permutation is applied, it will send the set of codewords to a set for which the message is again displayed in the first $k$ qubits.  It turns out that the answer to this question is yes; out of the
\be
\begin{pmatrix}
2^{n+c}\\
2^k
\end{pmatrix}\approx 2^{\left(n+c-k+\frac{1}{\log 2}\right)2^k}
\ee
possible sets of codewords there are $2^{(n+c-k)2^k}$ which have the message in the first $k$ bits.  Thus for a given initial string for the computer the probability that the permutation sends a generic set of codewords to a ``good'' set is
\be
P\approx e^{-2^k},
\ee
which for fixed $k$ we can easily beat by trying the various $2^c$ initial states for the computer.  It does require an $c\sim 2^k$-bit computer however.

\subsection{Implementing a General Unitary Transformation with Quantum Gates}\label{compsec}
\begin{figure}
\begin{center}
\includegraphics[height=4cm]{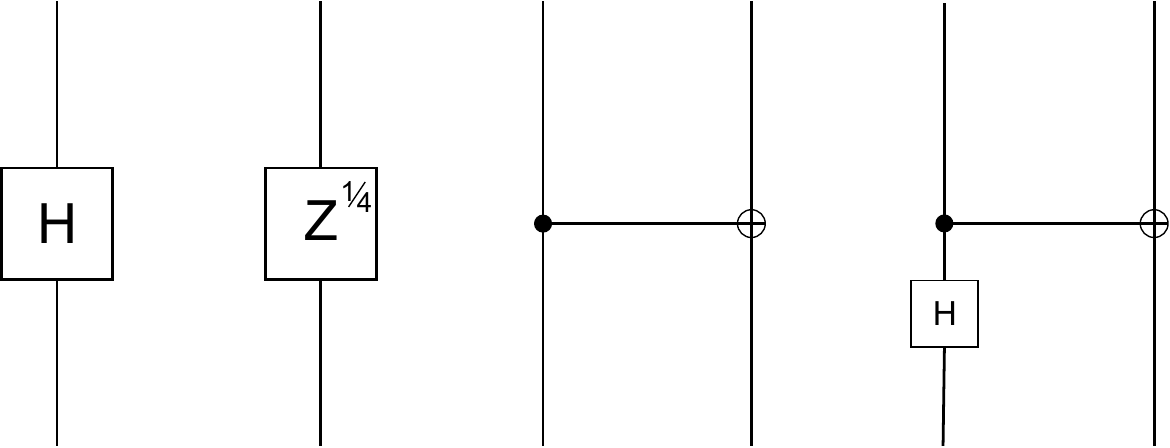}
\end{center}
\caption{The standard representations of the three gates described in the text, as well as a simple circuit that maps the product basis $|b_1,b_2\ran$ to a basis each element of which has the two qubits maximally entangled.  In the CNOT gate the addition is done at the hollow circle.}\label{gates}
\end{figure}

In our world, the result of the previous section, that doing a quantum computation at worst takes a time which is double exponential in the entropy of the computer, had better not be optimal if we are ever to do any quantum computation at all.  It can be improved upon of course, and the reason is that locality of interactions makes the dependence of $U_{comp}$ on time and computer size very special indeed.\footnote{This subsection is entirely pedagogical and contains no original material, good references are \cite{kitaev2002classical,preskillnotes,nielsen2010quantum}.}  Since no quantum computer has yet been built we do not know exactly how one might be implemented physically, but there is a widely accepted model for quantum computation called the \textit{quantum circuit model}.  In the quantum circuit model one imagines having a ``quantum memory'' consisting of $n$ qubits, on which one can easily act with some finite set of one- or two-qubit unitary transformations, called \textit{quantum gates}, on any qubit or pair of qubits.  The computer builds up larger unitary transformations by applying the various gates successively.  Interestingly the number of different types of gates needed to generate arbitrary unitary transformations with high precision is quite small. In fact, one is sufficient provided it is generic enough and that it can be applied to any two of the qubits (and in either order on those two).  A set of gates having this property is called \textit{universal}.  A specific set of three gates which is universal is the \textit{Hadamard} gate, which acts on a single qubit as 
\begin{align}\nonumber
H|0\ran=&\frac{1}{\sqrt{2}}\left(|0\ran+|1\ran\right)\\
H|1\ran=&\frac{1}{\sqrt{2}}\left(|0\ran-|1\ran\right),
\end{align}
the $Z^{1/4}$ gate which acts on a single qubit as\footnote{In this equation we see the unfortunate but standard convention in quantum computation theory that the Pauli-$z$ operator, usually written as $Z$, acts as $Z|0\ran=|0\ran$ and $Z|1\ran=-|1\ran$.  The Hadamard operator can then be interpreted as switching from the $Z$ eigenbasis to the $X$ eigenbasis.  As an example it is worth seeing to see how to build the Pauli operators $X$, $Y$, and $Z$ out of $H$ and $Z^{1/4}$ gates.}
\begin{align}\nonumber
Z^{1/4}|0\ran=&|0\ran\\
Z^{1/4}|1\ran=&e^{\frac{i\pi}{4}}|1\ran,
\end{align}
and the \textit{CNOT} gate $U_{cnot}$, for ``controlled not'', which acts on two qubits as
\be
U_{cnot}|b_1,b_2\ran=|b_1,b_1+b_2\ran
\ee
with the addition being mod $2$.  This gate flips the second bit if and only if the first bit is $1$.  There is a standard graphical notation for representing circuits, which we have already used in figure \ref{bigcomp}, and we illustrate it some more in figure \ref{gates}.

We can now ask how many gates are needed to make a complicated unitary transformation like $U_R$ in equation \eqref{compPsi}.  This is a good measure of the amount of time/space needed to actually do the computation, since we can imagine that the gates can be implemented one after another in a time that scales at most as a small power of $n$.  For a set of $f$ fundamental gates, the number of circuits we can make which use $T$ total gates is clearly 
\be
\left(\begin{pmatrix}
n\\
2
\end{pmatrix}f\right)^{T}\approx (n^2f)^{T}.
\ee

To proceed further we need some basic idea of size and distance for the unitary group.  The unitary group on $n$ qubits is a compact manifold of dimension $2^{2n}$, and we can parametrize its elements as
\be
U=e^{i\sum_{a=1}^{2^{2n}}c_a t^a}.
\ee
Here $t^a$ are generators of the Lie algebra of $U(2^n)$, and we can very roughly think of the $c_a$'s as parametrizing a unit cube in $\mathbb{R}^{2^{2n}}$.  Also roughly we can think of linear distance in this unit cube as a measure of distance between the unitaries.  For example say we wish to compute the difference between acting on some pure state $|\Psi\ran$ with two different unitary matrices $U_1$ and $U_2$ and then projecting onto some other state $\chi$:
\be
\lan \chi|(U_1-U_2)|\Psi\ran=\lan \chi|\left(I-U_2U_1^\dagger\right)U_1|\Psi\ran\approx -i\lan \chi|\sum_a \delta c_a t^a U_1|\Psi\ran.
\ee
If the sum of the squares of the $\delta c_a$'s is less than $\epsilon^2$, the right hand side will be at most some low order polynomial in $2^n$ times $\epsilon$.  This polynomial is irrelevant as we now see.  

Around each of our $(n^2f)^{T}$ circuits we can imagine a ball of radius $\epsilon$ in $\mathbb{R}^{2^{2n}}$.  The volume of all the balls together will be of order the full volume of the unitary group when
\be
(n^2 f)^{T}\epsilon^{2^{2n}}\approx 1.
\ee
Thus we see that in order to be able to make generic elements of $U(2^n)$ we need at least
\be
T\sim 2^{2n}\log\left(\frac{1}{\epsilon}\right)
\ee
gates, where we have kept only the leading dependence on $n$ and $\epsilon$.  As promised, because $\epsilon$ appears inside a logarithm the crude nature of our definition of distance has not mattered.  More importantly, we see that the number of gates is now only a single exponential in (twice) the entropy.  The fact that this crude counting bound can be achieved by an appropriate sequence of gates is known as the Solovay-Kitaev theorem. So the quantum circuit model is able to do arbitrary quantum computations much faster than our calculation of the previous section suggested; this is essentially because locality enables us to dynamically isolate parts of of the computer in such a way that we can push the chaos of the system into heating the environment (not explicitly modeled here) instead of messing up our computation.  

Given that we have so quickly beaten down a double exponential to a single exponential, one might be optimistic that further reduction in computing time is possible.  Unfortunately, in our universe that does not seem to be the case.  Simple modifications of the quantum circuit model such as changing the set of fundamental gates or considering higher spin fundamental objects instead of qubits, for example qutrits, make only small modifications to the analysis and don't change the main $2^{2n}$ scaling.  One could imagine trying to engineer gates that act on some finite fraction of the $n$ qubits all at once, perhaps by connecting them all together with wires or something, but it is easy to see that any such construction requires a number of wires exponential in $n$.  One could also try to parallelize by applying gates on non-overlapping qubits simultaneously whenever possible, as well as adding additional ``ancillary'' qubits.  As long as the number of extra qubits scales like some power of $n$, however, it is clear they cannot beat the $2^{2n}$. Even with exponentially many ancilla or wires just the travel time between the various parts of the computer will be exponential in $n$.  In the face of these difficulties the reader might be tempted to try using some sort of exotic nonlocal system like a black hole to do the computation, but this would just give up what the circuit model accomplished and most likely return us to even worse situation of the previous section.  In this paper we will adopt the widely held point of view that the quantum circuit model accurately describes what are physically realistic expectations for the power of a quantum computer.  

Thus if $U_{R}$ has no special structure, we do not expect Alice to be able to implement it (or its inverse) in time shorter than $2^{2n}$.  As discussed below equation \eqref{compPsi} however, $U_R$ was defined only by its action on a subspace of dimension $2^{k+m}$.  In studying the complexity of $U_R$ we are free to quotient $U(2^n)$ by multiplication on the right by a block diagonal matrix which acts as the identity on the subspace spanned by states of the form $|bh0\ran$.  The dimensionality of this quotient is $2^{k+m+n}$, and since it is good enough to find a circuit which produces any element of the coset containing $U_R$ the counting argument of this section suggests that Alice will be able to use a decoding circuit implementable in a time scaling like 
\be\label{dectime}
T\sim2^{k+m+n}.
\ee

\subsection{Why is Alice's Computation Slower than the Black Hole Dynamics?}\label{slowcomp}
We now turn to the question of whether or not the black hole dynamics constrain $U_R$ in any way that could help Alice implement it faster.  One thing we know about the black hole is that it produces the state \eqref{compPsi} relatively quickly, in a time that scales like $n^{3/2}$ for a Schwarzschild black hole.  This seems to suggest that Alice might be able implement $U_R^\dagger$ quickly by some sort of time-reversal.  This turns out not to be the case.  To explain this we introduce a slightly more detailed model of the dynamics that produce the state \eqref{compPsi}.  

To describe the evaporation process it is clearly necessary to have a Hilbert space in which the we can have black holes of different sizes.  We can write this as
\be
\mathcal{H}=\oplus_{n=0}^{n_f}\left(\mathcal{H}_{BH,n_f-n}\otimes \mathcal{H}_{R,n}\right).
\ee
Here the subscripts $n$ and $n_f-n$ indicate the number of qubits in the indicated Hilbert spaces.  The dimensionality of $\mathcal{H}$ is $n_f 2^{n_f}$.  We can imagine starting in the subspace with $n=0$ and then in each time-step acting with a unitary transformation that increases $n$ by one.  We will take the evolution on the radiation to be trivial.  The black hole becomes old after $n_f/2$ steps.  This ``adiabatic'' model of evaporation assumes equation \eqref{entguess} is exact and does not involve any energetics but, as discussed below equation \eqref{entguess}, it is not a bad approximation for the decay of a Schwarzschild black hole: the number of Hawking quanta produced is of order the entropy of the black hole, and so is their coarse-grained entropy.  

\begin{figure}
\begin{center}
\includegraphics[height=5cm]{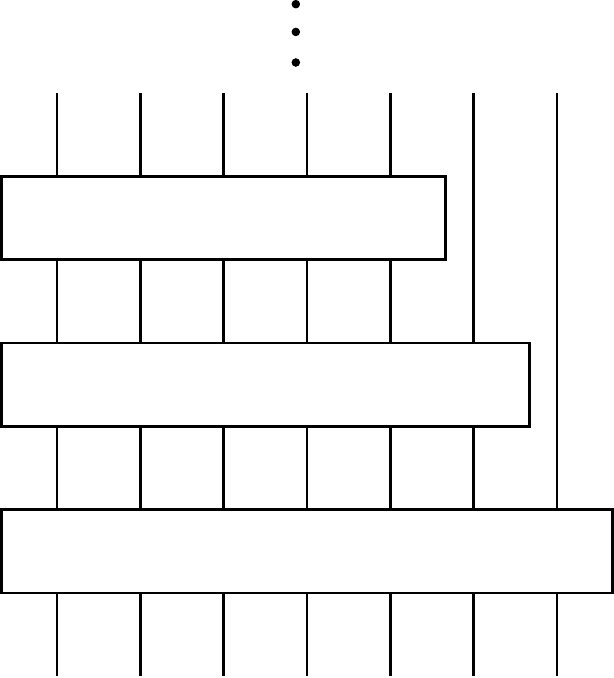}
\end{center}
\caption{The black hole dynamics for a 7-bit black hole.  With each step the subfactor we interpret as the radiation gets larger.}\label{encoder}
\end{figure}

An actual black hole formed in collapse will have some width in energy, which here means a width in $n$, but by ignoring this we can make a further simplification.  Starting in one of the $2^{n_f}$ states with $n=0$, the evolution never produces superpositions of different $n$.  So we can actually recast the whole dynamics as unitary evolution on a smaller Hilbert space of dimension $2^{n_f}$, but in which the interpretation of subfactors changes with time.  We illustrate this with a circuit diagram in figure \ref{encoder}.

With this simplification we can now combine all of the timesteps together into one big unitary matrix $U_{dyn}$ acting on our $2^{n_f}$ dimensional Hilbert space.  The matrix $U_R$ appearing in the state \eqref{compPsi}, which we now interpret as having been produced by $U_{dyn}$, will (unlike $U_{dyn}$) depend rather sensitively on the initial state, and since Alice only needs to be able to do the computation for some particular initial state we will for simplicity choose it to just have all the bits set to zero.  For $n>\frac{n_f}{2}$ we then expect
\be
U_{dyn}|00000\ran_{init}\approx\frac{1}{\sqrt{|B||H|}}\sum_{b,h}|b\ran_B|h\ran_HU_R|bh0\ran_R.
\ee

So this equation tells us something about $U_R$, whose complexity we are interested in understanding.  To proceed further we need to make some sort of assumption about $U_{dyn}$.  This is a question about the dynamics of quantum gravity so we can't say anything too precise, but for those black holes which are well understood in matrix theory \cite{Banks:1996vh} or AdS/CFT \cite{Maldacena:1997re,Witten:1998qj,Maldacena:2001kr} the dynamics are always some matrix quantum mechanics or matrix field theory.  Theories of this type can usually be simulated by polynomial-sized quantum circuits \cite{feynman1982simulating,lloyd1996universal,Abrams:1997gk,Jordan:2011ne}, so it seems quite reasonable to assume that $U_{dyn}$ can be generated by a polynomial number of gates.\footnote{Technically this also assumes that the mapping from the ``microscopic'' degrees of freedom on which the quantum mechanics looks simple to the ``macroscopic'' basis \eqref{Rbasis} is relatively simple.  For low energy fields outside the horizon this seems plausible to us, for example in AdS/CFT the well-known construction of \cite{Kabat:2011rz} seems to accomplish this in a straightforward way for operators outside the horizon.  We discuss this more in section \eqref{options} below.}  Such circuits are usually called ``small'', so more precisely we want to know the following: does the existence of a small circuit for $U_{dyn}$ imply the existence of a small circuit for $U_R$?  If the answer is yes, then our model would imply that Alice can decode $R_B$ out of the Hawking radiation fairly easily.  

It is clear that acting on the state $|00000\ran_{init}$ we can easily decompose $U_{dyn}$ into $U_R U_{mix}$, where $U_{mix}$ is a simple circuit that entangles the first four subfactors in $|00000\ran_{init}$:
\be\label{Umix}
U_{mix}|00000\ran_{init}=\frac{1}{\sqrt{|B||H|}}\sum_{b,h}|b\ran_B|h\ran_H|bh0\ran_R.
\ee
$U_{mix}$ is very easy to implement, we can just use the circuit on the right in figure \ref{gates} $n_f-n$ times for a total of $2(n_f-n)$ gates.  We can then define a new operator
\be
\tilde{U}_R=U_{dyn} U_{mix}^\dagger,
\ee
which has the property that
\be\label{tildeU}
\tilde{U}_R\frac{1}{\sqrt{|B||H|}}\sum_{b,h}|b\ran_B|h\ran_H|bh0\ran_R=\frac{1}{\sqrt{|B||H|}}\sum_{b,h}|b\ran_B|h\ran_HU_R|bh0\ran_R.
\ee
$\tilde{U}_R$ can obviously be implemented with a small circuit, and it apparently seems to be exactly what Alice needs; she can just apply the inverse circuit to the state \eqref{compPsi} and the decoding is accomplished.  Unfortunately for her this does not work.  Although the operator $\tilde{U}_R$ appears to only act on the radiation, the circuit this construction provides involves gates that act on all of the qubits.  While she is doing the decoding Alice does not have access to the qubits in $B$ and $H$, so she cannot directly use them.  Of course, if the circuit really acted as the identity operator on $B$ and $H$ for \textit{any} initial state this would not matter, she could just throw in some ancillary qubits in an arbitrary state to replace those in $B$ and $H$ and still use the $\tilde{U}_R^\dagger$ to undo $U_R$.  The problem is that \eqref{tildeU} holds \textit{only} when $\tilde{U}_R$ acts on the particular state $\frac{1}{\sqrt{|B||H|}}\sum_{b,h}|b\ran_B|h\ran_H|bh0\ran_R$.  This can be traced back to the fact that the definition of $U_R$ in the first place depended on the initial state of the black hole on which $U_{dyn}$ acts.  

Although Alice cannot use these small circuits to decode the entanglement, she can move it around.  For example acting with a CNOT gate three times on two qubits, switching which qubit is the ``control'' qubit each time, exchanges the pair:
\be
|b_1,b_2\ran\to|b_1,b_1+b_2\ran\to|b_2,b_1+b_2\ran\to|b_2,b_1\ran,
\ee
so by adjoining a set of $n$ ancillary qubits to the state $|\Psi\ran$, Alice can use this operation to achieve
\be
\frac{1}{\sqrt{|B||H|}}\sum_{b,h}|b\ran_B|h\ran_HU_R|bh0\ran_R|000\ran_{anc}\to \frac{1}{\sqrt{|B||H|}}\sum_{b,h}|b\ran_B|h\ran_H|000\ran_RU_R|bh0\ran_{anc}.
\ee
This trivializes the state of the radiation, but of course doesn't really accomplish much since testing the entanglement still requires undoing $U_R$.  It does show however that Alice can move the quantum information from the radiation to a more ``stable''  quantum memory in a short amount of time.  

The lesson of this section is that because Alice does not have access to all of the qubits in the system, she is unable to simply time-reverse the black hole dynamics and extract $R_B$ in a time that is polynomial in the entropy.  Without such a simple construction, she will in general be left with no option but to brute-force her construction of $U_R^\dagger$ using of order $2^{n+k+m}$ gates.\footnote{Of course even if she could just time reverse the black hole, the circuit would still take of order the evaporation time to run.  In fact since the black hole is old it would probably take longer to run than the evaporation.  We are not comfortable with this argument as a way out of firewalls however.  Often when there is a general algorithm that gives a polynomial circuit to do something, special details of the problem and tricks like parallelization can be exploited to get polynomial increases in the running speed.}  It is still possible that some yet-unknown special features of black hole dynamics will conspire to provide a simple circuit for $U_R$, but it would be rather surprising.  After all there are many ways a unitary could be atypical, and most still require exponentially many gates.  In the following section we will see that for some analogous questions in the theory of error-correcting codes, within the context of simple circuits with $O(n^2)$ gates it is possible to run into problems which almost certainly take exponential time to solve.  These results will unfortunately not be directly applicable here, but by recasting Alice's decoding task as an error correction problem we will be able to get more intuition about how the exponential arises.  We will also see that being able to efficiently perform a decoding very similar to $U_R$ would have very unlikely implications for the complexity class {Quantum Statistical Zero-Knowledge}.

\subsection{Two Counting Arguments}
So far we have mostly restricted ourselves to thinking about a particular black hole microstate, which in the previous section we took to be an element of the basis where $U_{dyn}$ can be constructed out of a polynomial-size quantum circuit.  By considering more general initial states we can prove that for a generic initial state the decoding necessarily takes exponential time.  Restating the setup of the previous section, we can think of the black hole as a polynomial-size circuit $U_{dyn}$ which acts on a set of microstates $|i\ran$ as
\be
U_{dyn}|i\ran=\frac{1}{\sqrt{|B||H|}}\sum_{bh}|b\ran_B|h\ran_H U_R(i)|bh0\ran_R.
\ee
The point here is that we must be able to recover any initial state from the final state, so the unitaries $U_R(i)$ must all be distinct elements of the coset $U(2^n)/U(2^n-2^{m+k})$.  But how many initial states are there? If we ignore the coarse-graining that made equation \eqref{entguess} only approximate, then there is a basis of size $2^{n+k+m}$ for the total number of initial states of the black hole.  The full set of initial states includes superpositions however, and we can count them by constructing an $\epsilon$-net like that discussed around equation \eqref{norms}.  The total number of initial states then is roughly $\left(\frac{1}{\epsilon}\right)^{2^{n+k+m}}$.  The point then is that this is comparable to the total number of elements in the coset, which also scales like $\left(\frac{1}{\epsilon}\right)^{2^{n+k+m}}$.  Since each state $|i\ran$ must have its own distinct $U_R(i)$, the vast majority of initial microstates must be decoded by generic $U_R$'s which necessarily take time of order $2^{n+k+m}$ to implement.  Being more careful about the coarse-graining decreases the number of initial states somewhat, but the number is still much too large for anything more than a vanishingly small fraction to be decodable in polynomial time.  

Proponents of the AMPS experiment could still hope for a basis of initial states where each basis element is easy to decode.  A natural basis to try is the one in which $U_{dyn}$ is polynomial size, but we will present arguments in the following section that even in this basis the decoding is hard.  One could however try to construct such a basis by taking a basis of final states where the decoding is trivial and then evolving it back with $U_{dyn}^\dagger$.\footnote{This possibility was suggested to us by Raphael Bousso.}  This construction can be disrupted however by taking into account the coarse-graining in equation \eqref{entguess} more carefully.  More explicitly, let's say that $n+k+m=S_{initial}/\log 2+\Delta n$ for some $\Delta n > 0$.  The number of initial states is then double exponential in $n+k+m-\Delta n$.  It is then not hard to see that even when $\Delta n=1$, this set of initial states produces a set of $U_R$'s which form a doubly-exponential small fraction of the coset $U(2^n)/U(2^n-2^{m+k})$.  If each $U_R$ were chosen at random and independently, it is doubly exponentially unlikely that even one of these states would produce a $U_R$ which can be implemented with a polynomial-size circuit.  The initial states constructed by time reversal of trivial decoding states would not correspond to black holes; they would be states in some larger Hilbert space involving entanglement between the black hole and the radiation field.  
  
\section{Quantum Coding and Error Correction}\label{decsect}
The theory of quantum error correcting codes has interesting implications for the AMPS experiment, which we discuss in this section.\footnote{A recent paper \cite{Verlinde:2012cy} also discussed the AMPS argument in the language of error correction; their use of error correction was quite different from that we discuss here.}  We will review the main points of this theory assuming no prior experience with the subject.  This will be something of a sidetrip from our main exposition, so casual readers may want to skip over this section in a first reading.  For the impatient the conclusions of relevance for AMPS are summarized at the end of the section.  

\subsection{Review of Error Correcting Codes}\label{ecrev}
\begin{figure}
\begin{center}
\includegraphics[height=5cm]{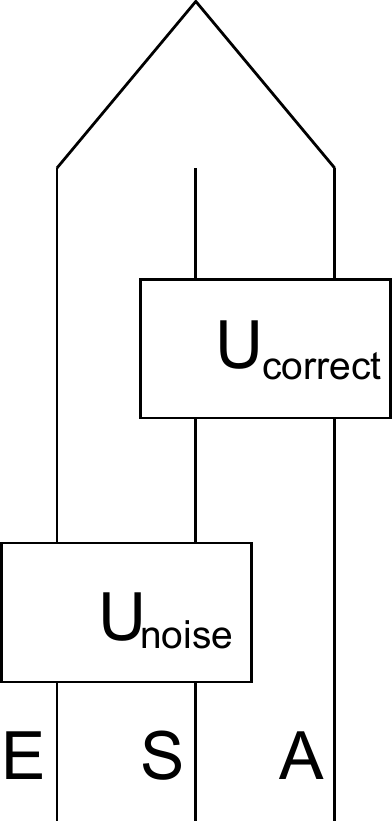}
\end{center}
\caption{Quantum error correction.  Here $S$ is the system whose state we want to restore, $E$ is the environment it becomes entangled with via the interaction $U_{noise}$, and $A$ is the ancilla it interacts with via $U_{correct}$.  At the end the environment is entangled with the ancilla and the system $S$ is in the same quantum state it started in.}\label{correction}
\end{figure}

Typically quantum systems cannot be isolated from their environment.  It is interesting to understand to what extent a system which began in an unknown state can be restored to that state after it has interacted nontrivially with its environment.  Usually this is done by introducing another ancillary system and transferring the entanglement with the environment from the system whose state we want to restore to the ancillary system.  We show this pictorially in figure \ref{correction}.

Error correction is not always possible.  For example, say that the transformation $U_{noise}$ from figure \ref{correction} is such that it results in the system $S$ being maximally entangled with the environment $E$.  There is no information about its initial state remaining in $S$, and no choice of $U_{correct}$ will allow recovery.  What is perhaps surprising is that it is ever possible to restore the initial state after a nontrivial $U_{noise}$ has acted.  To see that this can be done, following Shor \cite{shor1995scheme} we consider an arbitrary superposition $|\Psi\ran=a_+|+\ran+a_-|-\ran$ of the following two nine-qubit states:
\be
|\pm\ran\equiv \frac{1}{2^{3/2}}\left(|000\ran\pm|111\ran\right)\left(|000\ran\pm|111\ran\right)\left(|000\ran\pm|111\ran\right).
\ee
As a crude model of interaction with the environment, we can imagine that the state $|\Psi\ran$ will be acted on randomly by a single Pauli operator $X$, $Y$, or $Z$ on one of its nine qubits.  After this can we restore the state $|\Psi\ran$ without destroying it?  As Shor explained in \cite{shor1995scheme}, we can.  The idea is to measure the following set of eight ``check'' operators: 
\begin{align}\nonumber
Z_1 Z_2,& \qquad Z_2 Z_3,\qquad Z_4Z_5,\qquad Z_5Z_6,\qquad Z_7Z_8,\qquad Z_8 Z_9,\\
&X_1X_2X_3X_4X_5X_6,\qquad \mathrm{and}\qquad X_4X_5X_6X_7X_8X_9.
\end{align}
Since the states $|\pm\ran$ are both eigenstates of eigenvalue one for all eight of these operators, this measurement will do nothing to the state $\Psi$.  Let's say however that interaction with the environment has caused the state to be acted on by $X_1$.  This commutes with the last seven check operators, so their eigenvalues are unaffected, but it will change the result of measuring $Z_1Z_2$ to from $1$ to $-1$ for both states $|\pm\ran$.  It is easy to see that none of the other possible single-Pauli errors will give this signature as a result of measuring the check operators.  We can then ``repair'' the error up to an overall irrelevant phase by acting with $X_1$.  Similarly say that the error acts with $Z_1$.  This will now flip the eigenvalue of $X_1X_2X_3X_4X_5X_6$ without affecting any of the other check operators.  There are now two other single-Pauli errors with the same signature, $Z_2$ and $Z_3$, but we can correct any of the three up to an overall phase by acting with $Z_1$ on the state.  In this manner it is easy to see that any single-Pauli error can be corrected.\footnote{It is not hard to describe this procedure in terms of unitary interaction $U_{correct}$ with an ancillary system.  For example to measure $Z_1Z_2$ we introduce a single ancilla qubit in the state $|0\ran_{anc}$ and then use two CNOT gates with the first and second qubits being the control bits in the gate.  This accomplishes $|b_1,b_2\ran|0\ran_{anc}\to|b_1,b_2\ran|b_1\ran_{anc}\to|b_1,b_2\ran|b_1+b_2\ran_{anc}$, which writes the result of the measurement onto the ancillary qubit.}  This protocol is called the Shor code, and it was the first quantum error correcting code to be discovered.

The Shor code works for two reasons.  One is that it is \textit{redundant}, meaning that the number of bits of information it protects is significantly fewer than the number of physical bits present.  This allows ``room'' for noise to creep in without disrupting the message.  The other reason is that it is nonlocal; the information is carried in the entanglement between multiple qubits, which protects it against local decoherence and depolarization.  These observations motivate the general definition of a quantum code as a \textit{code subspace} - a $k$-qubit subspace of a larger $n$-qubit Hilbert space out of which we can build states we wish to protect.  In the Shor code the code subspace is spanned by the states $|\pm\ran$.  Given a code subspace $\mathcal{H}_C\subseteq\mathcal{H}$ we can always define an encoding transformation $U_{enc}$ with the property that
\be\label{encoding}
|\bar{c}\ran=U_{enc}|c_1\ldots c_k,\underbrace{0\ldots 0}_{n-k\text{ zeros}}\ran,
\ee
with $|\bar{c}\ran$ a complete basis for $\mathcal{H}_C$.  

To describe error correction more generally, we first need to say more about how to understand the noise generated in a quantum system $S$ interacting with an environment.  In such a situation we can always write
\be \label{eqnUnoise}
U_{noise}|s0\ran=\sum_{s',e}\lan s'e|U_{noise}|s0\ran|s'e\ran\equiv \sum_e M_e|se\ran,
\ee
where $|e\ran$ is an orthonormal basis for the environment and the operators $M_e$ are called Kraus operators \cite{kraus1983states}.  They act only on the system $S$, have matrix elements 
\be
\lan s'|M_e|s\ran=\lan s'e|U_{noise}|s0\ran, 
\ee
and obey
\be\label{krausorth}
\sum_e M_e^\dagger M_e=1.
\ee
A set of operators $M_e$ obeying \eqref{krausorth} is sometimes called a \textit{quantum channel}, or occasionally a \textit{superoperator}.

The Kraus operators can be awkward to work with in practice because their definition depends on the details of the interaction with the environment, which we usually do not know.  It is therefore convenient to expand them in a standard basis $E_\alpha$ of operators such as the $2^{2n}$ distinct products of the Pauli operators
\be
M_e=\sum_\alpha C_{\alpha e} E_\alpha,
\ee
and rewrite
\be\label{Eerr}
U_{noise}|s\ran|0\ran=\sum_\alpha E_\alpha|s\ran|\alpha\ran.
\ee
Here we have defined $|\alpha\ran\equiv \sum_e C_{\alpha e}|e\ran$, which are no longer necessarily orthonormal.  

In this language there is a necessary and sufficient condition for when exact error correction is possible for a given code \cite{bennett1996mixed,knill1997theory}.  A set $\mathcal{E}$ of errors $E_\alpha$ is exactly correctable if and only if
\be\label{correctcond}
\lan\bar{c}'|E_\alpha^\dagger E_\beta|\bar{c}\ran=\delta_{cc'}C_{\alpha\beta}
\ee
for any $|\bar{c}\ran$, $|\bar{c}'\ran$ in some orthonormal basis for the code space, for any $E_\alpha,E_\beta\in \mathcal{E}$, and with the coefficents $C_{\alpha\beta}$ independent of $c$.  If the Kraus operators produced by the interaction with the environment can be written only in terms of $E_\alpha$'s in a set $\mathcal{E}$ with this property, then a generalization of the procedure described above for the Shor code will always allow the state to be recovered perfectly.

To build some intuition for this criterion, consider the \textit{trivial code} $U_{enc}=1$.  The codespace is spanned by states of the form
\be\label{trivcode}
|\bar{c}\ran=|c_1\ldots c_k, 0\ldots 0\ran.
\ee
There is an obvious set of $n-k$ ``check'' operators acting on this code space with eigenvalue one; the spin $Z$ operators acting on the last $n-k$ qubits.  As suggested above we can take the $E_\alpha$'s to be some subset of the $2^{2n}$ distinct products of Pauli matrices, which in general we can write as
\be\label{PauliProd}
X_1^{\alpha_1}Z_1^{\beta_1}\ldots X_n^{\alpha_n} Z_n^{\beta_n}.
\ee
The parameters $\alpha_1, \beta_1, \ldots$ take the values $0$ or $1$.  This set of errors has the special property that each element either commutes or anticommutes with each of the check operators, so the condition \eqref{correctcond} implies that the matrix $C_{\alpha\beta}$ will be zero for any two errors $E_\alpha$, $E_\beta$ in $\mathcal{E}$ unless they commute with the same set of check operators.  Moreover it further implies that in the event that $C_{\alpha,\beta}\neq 0$ then the product $E_\alpha^\dagger E_\beta$ is equal to a product of check operators.  Now say we have a channel
\be\label{errorstate}
\sum_\alpha E_\alpha|\bar{c}\ran|\alpha\ran,
\ee 
with all $E_\alpha$ with $|\alpha\ran\neq0$ in a set $\mathcal{E}$ obeying \eqref{correctcond}.  Channels with this property are called \textit{exactly correctable}.  By measuring the check operators we can collapse this state to one where all remaining $E_\alpha$'s commute with the same with the same set of check operators.  We may then pick any one of the remaining $E_\alpha$'s and apply $E_\alpha^\dagger$ to restore the state $|\bar{c}\ran$, since in each term this produces some product of check operators which acts trivially on the code space.  This correction procedure is easily extended to the general case of nontrivial $U_{enc}$.  We can simply take the set $\mathcal{E}$ of correctable errors for the trivial code and conjugate it by $U_{enc}$ to define $\mathcal{E'}=U_{enc}\mathcal{E}U_{enc}^\dagger$.  The check operators can be taken to be $U_{enc}Z_i U_{enc}^\dagger$ for the last $n-k$ Pauli spin Z operators.  This construction is not unique since for a given code subspace there will be many $U_{enc}$'s which satisfy \eqref{encoding}.  Of course channels are not always exactly correctable, and when they are not a would-be quantum repair-person needs to look at the set of possible corrections consistent with the result of the check operator measurements and then make her best guess about which one to apply based on her understanding of which errors are most likely.  This procedure is sometimes called \textit{Maximal Likelihood Decoding}.

The correction protocol just described is not necessarily the best one in general situations.  The transformation $U_{enc}$ relates the ``logical'' basis in which the code is trivial to the ``computational'' basis in which the interactions with the environment are simple.  The set of conjugated Pauli errors we just discussed is simple from the point of view of the logical basis, but depending on the complexity of $U_{enc}$ it could be quite unnatural from the point of view of the computational basis.  In particular we usually expect the Kraus operators to be simple in the computational basis, and insisting on expanding them in the conjugated Pauli basis will lead to more nonzero $|\alpha\ran$'s than are really necessary.  A simple example of this that we will use below is the \textit{depolarizing channel}, which acts on a single qubit coupled to a $4$-state environment as
\be
U_{noise}:|\Psi\ran |0\ran\longmapsto \frac{1}{2}\Big[|\Psi\ran|0\ran+X|\Psi\ran|1\ran+Y|\Psi\ran|2\ran+Z|\Psi\ran|3\ran\Big].
\ee
The Kraus operators for this channel are $I/2,X/2,Y/2,Z/2$, and it is not hard to verify that it maximally entangles the qubit with the environment, erasing any information about the initial state $|\Psi\ran$.  More generally, the \textit{erasure channel} is defined as applying this channel to some known set of $l$ out of our $n$ total qubits.  The total set of errors appearing in the channel is then $2^{2l}$, but if we conjugate them by $U_{enc}$ and re-expand in the Pauli basis we in general find more.  Moreover even if the set of $2^{2l}$ erasures obey the correctability condition \eqref{correctcond}, it will not in general be possible to find some set of conjugated Pauli's that do.  For \textit{stabilizer codes} \cite{gottesman1997stabilizer}, a widely studied set of codes that we review briefly in the following subsection, the encoding transformation sends Pauli errors to Pauli errors and we can use the check operator correction procedure described above.  For more general $U_{enc}$ we need to do something else.\footnote{Daniel Gottesman has suggested a simple ``something else'' to us, in the case where the matrix $C_{\alpha\beta}$ is diagonal.  In that case the errors $E_\alpha$ in $\mathcal{E}$ send the code space to a set of mutually orthogonal subspaces of the Hilbert space; this will typically be the case in the erasure channel if $2l<n-k$.  To do the correction we can just measure the projection operators onto each of these subspaces in succession, stopping when the measurement returns one instead of zero.  We can then just apply the correction appropriate for that subspace.  This procedure clearly takes an amount of time equal to the number of possible errors, which in the erasure channel is $2^{2l}$.}   

\subsection{Computational Complexity, Stabilizer Codes, and \NP-Hardness}\label{corrcomp}
One obvious source of computational hardness in the error correction procedure described in the previous section is that in measuring the check operators and applying corrections, we need to repeatedly use the encoding transformation $U_{enc}$.  If this unitary is difficult to implement as a quantum circuit, the error correction procedure will be very time-consuming.  In our discussion of the AMPS experiment below however, $U_{enc}$ will always be implementable with a small circuit.  The rest of the correction procedure seems straightforward, but there is one crucial point where after measuring the check operators we need to pick a correction $E_\alpha^\dagger$ to apply.  In general there are $2^{2n}$ possible errors, out of which $2^{n+k}$ are consistent with any given result for the check operator measurements.  For a general channel it might be quite hard to decide which of these $2^{n+k}$ corrections to apply.  Since there are $2^{n-k}$ possible results for measuring the check operators, it is impractical to carry around a list of which correction to apply for each syndrome.  Without such a list however, it will generally take exponential time in $n$ to determine which check operator to apply since one will just have to go through the list of $2^{n+k}$ potential corrections and compute the probability for each that it is the correct one to apply.  In this section we sketch some known results about the hardness of this step for a widely studied set of quantum error correcting codes, the stabilizer codes.

Stabilizer codes \cite{gottesman1997stabilizer} are defined by the property that the $n-k$ check operators in the computational basis are all just products of Pauli operators of the form \eqref{PauliProd}.  Clearly both the Shor code and the trivial code are stabilizer codes.  The reason for the name is that the check operators generate a $2^{n-k}$-element abelian subgroup $\mathcal{S}$ of the full group $\mathcal{G}$ of $2^{2n}$ products of Pauli operators, with the property that with respect to the action of $\mathcal{G}$ on the $n$-qubit Hilbert space $\mathcal{S}$ is the stabilizer subgroup of the $k$-qubit code subspace.  Thus rather than giving the codespace explicitly we can instead define it by picking some Abelian subgroup of the Pauli group; in practice this is a much more convenient way of defining an error correcting code.  Stabilizer codes are fairly easy to encode: for any stabilizer code there is a relatively simple construction \cite{gottesman1997stabilizer} of a circuit of size $O(n^2)$ which implements $U_{enc}$ exactly.  

Stabilizer codes have the special property that we can expand the errors in Pauli products of the form \eqref{PauliProd} in the computational basis instead of the logical basis and still have each error commute or anticommute with each check operator.  This enables an elegant way of describing the relation of the errors to the results of the check operator measurements.  For each check operator we can define a $2n$ component row vector where the first $n$ components are its $\alpha$'s from the parametrization \eqref{PauliProd} and the last $n$ are its $\beta$'s.   We can then assemble these row vectors into an $(n-k)\times 2n$ matrix $H$ which is often called the parity check matrix.  For each error we can make a $2n$-component column vector $e$ whose first $n$ elements are the $\beta$'s for the error and whose last $n$ components are its $\alpha$'s.  Finally we can define the \textit{syndrome} $s$ of the error as an $n-k$ component column vector whose $i$th element is $0$ if the error commutes with the $ith$ check operator and is $1$ if it anticommutes.  It is then not hard to check that $He=s$, with the matrix multiplication done in the field $\mathbb{Z}_2$.  Thus the set of errors $e$ consistent with the syndrome $s$ are just the solutions of this linear equation.  

One simple channel in which one could study the hardness of stabilizer error correction is the erasure channel defined in the previous subsection.  In this case the possible error vectors $e$ are nonzero only for a known set of $l$ out of the $n$ qubits.  We are then free to remove the $2(n-l)$ columns from $H$ which correspond to qubits which are never erased, finding a new matrix $H'$ which is only $(n-k)\times 2l$.  When the inequality $2l<n-k$ is satisfied the matrix equation $H'e'=s$ is overdetermined and can always be solved by Gaussian elimination in polynomial time unless no solution exists.  Indeed this inequality characterizes the \textit{channel capacity} of the erasure channel, since it can be shown that when it is violated it is impossible for the conditions \eqref{correctcond} to be satisfied, while when it is satisfied \eqref{correctcond} will also be satisfied for a typical stabilizer code \cite{preskillnotes}.  Thus stabilizer codes in the erasure channel can be corrected in polynomial time.  We emphasize however that this matrix structure is special to stabilizer codes, and that for more general codes with polynomial size $U_{enc}$ this procedure will not work.  The procedure described in the footnote at the end of the last section will work, but it requires a time $2^{2l}$ which will be quite large when $l$ is order $n$.

This result may not seem so promising from the point of view of the goal of this article, but in fact there is a slightly more complicated channel in which it is possible to show that stabilizer decoding is \NP-\textit{complete} \cite{berlekamp1978inherent,hsieh2011np}.  This is the gold standard in evidence that a classical computation cannot be performed in polynomial time. Good reviews of the basic properties of \NP-completeness and its possible relevance for physics are \cite{Denef:2006ad,Aaronson:2005qu}.  Unfortunately the technical details of this argument restrict it to showing \NP-completeness \textit{only} if $k$ grows nontrivially with $n$, since the exponential search which arises involves a search through $2^k$ things.  This is an assumption we'd like to drop in the context of Alice's experiment, since we'd like her to be unable to verify the entanglement even when $k$ is order one, but the result is still indicative of our main point since it shows quite convincingly how exponential time may be required for error correction even when $U_{noise}$ and $U_{enc}$ are both polynomial size.\footnote{In more detail, the argument of \cite{hsieh2011np} splits the error correction into two parts, one of which is proven to be \NP-complete for large $k$, and thus likely to require time exponential in $k$.  The other part appears to require exponential time in $n$, but they do not study its complexity in detail.  We expect stabilizer decoding in general to be hard even at fixed $k$ and large $n$.}  The channel studied by \cite{hsieh2011np} is one in which $X$ or $Z$ errors can occur on any qubit with probability $p$.  The probability for a general error $E_{\alpha}$ is 
\be
Pr(E_\alpha)=p^{w_{\alpha}}(1-p)^{2n-w_{\alpha}},
\ee
where $w_{\alpha}$ is called the \textit{weight} of the error and is defined as as the number of $\alpha_i$'s and $\beta_i$'s in the parametrization \eqref{PauliProd} which are nonzero.\footnote{That $Y$ is special is an undesirable asymmetry that is necessary for the proof of \cite{hsieh2011np}.  This seems to be a technical problem, and we expect that correcting stabilizer codes is still \NP-complete for the ``depolarizing channel'', where the probability for a $Y$ occuring is also $p$.  It seems unlikely that making $Y$ errors more likely would make the decoding easier but at the moment there is no proof.}  In order to correct this channel one needs to find the most likely set of errors, all fixable by a single correction, which is consistent with the syndrome $s$.  As explained in \cite{hsieh2011np}, the ability to answer this question for general stabilizer codes in polynomial time in $n$ would allow polynomial time in $k$ solution of the following question about matrices:
\begin{itemize}
\item Say we are given an integer $w>0$, an $(n-k)\times n$ matrix $A$, and a vector $y$ with $n-k$ components, with the components of the latter two being elements of the finite field $\mathbb{Z}_2$. Is there a vector $x$ with $n$ components in $\mathbb{Z}_2$, of which at most $w$ are nonzero, with the property that $Ax=y$?
\end{itemize}
This problem is called the \textcs{Coset Weights} problem; it appears in classical coding theory and was long ago shown to be \NP-complete \cite{berlekamp1978inherent}.  Since \textcs{Coset Weights} is \NP-complete, this then would allow polynomial time solution of any size $k$ problem in the computational class \NP.  This is still not known to be impossible, but the very widely believed conjecture $\textcs{P} \neq \NP$ precludes it.  For a discussion of the vast support for this conjecture, see for example \cite{moore2011nature}.

\subsection{Alice's Task as Error Correction}
Having reviewed all this formalism we now see what it has to tell us about Alice's computing task.  There is a straightforward way to recast what Alice is trying to do as a quantum coding problem \cite{HaydenPreskill}.  We can write the state \eqref{compPsi} of the black hole as
\be
|\Psi\ran=\frac{1}{\sqrt{|B|}}\sum |b\ran_B |\bar{b}\ran,
\ee
where 
\be\label{bhcode}
|\bar{b}\ran \equiv \frac{1}{\sqrt{|H|}} |h\ran_H U_R |bh0\ran_R
\ee
is a basis for a $k$ dimensional subspace of $\HH\otimes\HR$.  We can obviously interpret this subspace as a quantum code, with encoding transformation
\be\label{bhenc}
U_{enc}\equiv U_{R}U_{mix,H}.
\ee
Here $U_{mix,H}$ is a simple entangling transformation analogous to $U_{mix}$ from equation \eqref{Umix}, but here entangling only  the $m$ qubits of $H$ and the $n+1$ to $n+m$th qubits of $R$.  For later convenience we will take $U_{mix,H}$ to act on each pair of qubits as
\be\label{entangler2}
|b_1b_2\ran\mapsto \frac{1}{\sqrt{2}}\left((-1)^{b_1}|b_1b_2\ran+|b_1+1,b_2+1\ran\right),
\ee
which is not quite the transformation given by the circuit in figure \ref{gates}; to implement it we need to act with an extra CNOT gate prior to the Hadamard transformation and CNOT gates in figure \ref{gates}.  In this language, we can think of Alice not having access to $H$ as putting this code through an erasure channel that erases all qubits in $H$.  

This simple recasting unfortunately does not allow us to learn anything about the computational complexity of $U_R$, since the state $|\Psi\ran$ has the special property that it can be decoded just by acting with $U_R^\dagger$.  The error correction is no harder than the encoding, whose complexity we don't know a priori.  We can do better by recalling the discussion of section \ref{slowcomp}, where we interpreted the state $|\Psi\ran$ as arising from the action of a polynomial size circuit $U_{dyn}$
\be
U_{dyn}|0\ran_{BHR}=\frac{1}{\sqrt{|B||H|}}\sum_{bh}|b\ran_B |h\ran_HU_R|bh0\ran_R.
\ee
We'd like to interpret $U_{dyn}$ as an encoding transformation and $U_R$ as a correction operation, but this is not quite manifest here since $U_{dyn}$ always acts on the same state so the code space appears one dimensional whereas we'd like it to be $2^k$ dimensional.  We can fix this by introducing an additional system $B'$, with the same number of qubits as $B$, and then entangle it with $B$ by acting with a transformation $U_{mix,B'}$ which acts as \eqref{entangler2} on the $ith$ qubits of $B'$ and $B$ for all $i\in 1,\ldots,k$.  We can then define a $k$-qubit code subspace of this $2k+m+n$ qubit Hilbert space via the encoding transformation $U_{enc}\equiv U_{mix,B'}U_{dyn}$ such that
\be\label{Bpstate}
U_{enc}|b'\ran_{B'}|0\ran_{BHR}=\frac{1}{\sqrt{|B||H|}}\sum_{bh}U_{mix,B'}|b'b\ran_{B'B} |h\ran_HU_R|bh0\ran_R.
\ee
The errors introduced by the environment we again take to be erasures, now acting on both $B$ and $H$.  With $B$ and $H$ erased, the only way to extract $b'$ and restore the initial state is to unscramble the entanglement between $B'$ and $R$, which effectively requires being able to implement $U_R^\dagger$.  More concretely, say that we have a polynomial size circuit for $U_R$.  We may use it to act on the state \eqref{Bpstate} with $U_R^\dagger$.  We may then act with $U_{mix,B'}$, now with the second element of each pair of qubits taken from the first $k$ qubits of $R$ instead of from $B$, which we no longer have access to.  This produces a state
\be
|b'\ran\frac{1}{\sqrt{|B||H|}}\sum_{bh}|b\ran_B|h\ran_H|bh0\ran_R,
\ee 
so we have recovered the information $b'$.  From here we can explicitly restore the initial state in polynomial time by using ancilla to return $BHR$ to the state $|0\ran_{BHR}$ and then using $U_{enc}$ to get back to the state \eqref{Bpstate} as desired.  
We may thus apply the lesson of the previous section, which is that in general doing error correction for codes with small encoding circuits requires exponential sized circuits.  For stabilizer codes in the erasure channel we saw that this was not true, but there is no reason to expect that the encoding map \eqref{Bpstate} produces a stabilizer code.  For more general $U_{enc}$ the only known correction procedure even in the erasure channel is the one described in the footnote at the end of section \ref{ecrev}, which takes a time $2^{2l}$ where $l$ is the number of erased qubits.\footnote{By reshuffling factors between $U_{enc}$ and $U_{noise}$ we can turn erasures for a general code into a more general channel for a stabilizer code, which for simplicity we can just take to be the trivial code.  The \NP-completeness results of \cite{hsieh2011np} strongly suggest that no general polynomial time error correction algorithm exists for putting stabilizer codes through general channels, and essentially prove it when $k$ is large provided that $\textcs{P}\neq \NP$.}\footnote{It is interesting to note that since we here have $l=k+m$, the correction time $2^{2(k+m)}$ is actually a bit faster than our estimate \eqref{dectime} which came from counting gates.  This is because the correction procedure from the footnote just mentioned involves doing quantum operations on $B'$, which is fictitious and not available from the point of view of the original black hole problem.  In other words, any circuit for $U_R$ can be used to correct the channel just constructed but the converse is not true; a generic correction procedure for this channel cannot be converted into a circuit for $U_R$ of comparable complexity.}

So far we have only discussed errors arising from Alice not having access to $B$ and $H$.  There are of course other errors that can occur, having to do with the practical difficulty of controlling the Hawking radiation.  In particular an order one fraction of the radiation will be gravitons, which are very hard to detect at all, never mind coherently manipulate.  It was shown long ago by Page \cite{pageevap} that the rate of Hawking radiation into a field of spin $s$ decreases as $s$ increases, so for Schwarzschild black holes in our universe the fraction will be small. The question of how many bits can be lost without losing the ability to error correct with high probability of success has been studied in the literature, and as discussed in the previous subsection for the erasure channel the number of bits which are lost must be less than $\frac{n-k}{2}$. For a typical stabilizer code any smaller number of erasures can be corrected with high probability of success \cite{preskillnotes}.  This will typically be true for more general codes like the one discussed here as well since they should protect information at least as well as stabilizer codes.  For the channel just constructed, what was called $n$ in our general discussion of error correction is $2k+m+n$ in terms of the parameters of our black hole model, so we can lose at most $\frac{k+m+n}{2}$ bits.  $k+m$ have already been lost since Alice does not have access to $B$ and $H$, so she can lose $\frac{n-k-m}{2}$ more and still be able to error correct.  Say that the fraction of the radiation which is gravitons is $\alpha$.  Since $\alpha<\frac{1}{2}$, this means that if Alice waits long enough she can lose \textit{all} of the gravitons and still be able extract the entanglement accurately with room to spare for correcting additional errors.  Of course doing this error correction can be an additional source of computational hardness, which will only make it even harder to complete the decoding before the black hole evaporates.  This discussion shows however that without taking into account computational complexity, the difficulty of measuring gravitons does not prevent the AMPS experiment from being done.


\subsection{Error Correction and Zero-Knowledge Proofs}\label{qszksubsect}

We've seen that the known \NP-hardness results about decoding quantum error correcting codes, although suggestive, break down for technical reasons when $k$ is order one and thus unfortunately can't be invoked directly to draw any firm conclusions about $U_R$ in that case.  Moreover one might worry that the appearance of the \NP-complete problem \textcs{Coset Weights} in the argument of \cite{hsieh2011np} was a consequence of their particular decoding strategy; perhaps some more exotic correction procedure could somehow do the error correction without having an intermediate step that allowed solution of \textcs{Coset Weights}.  We find this unlikely, but both of these concerns can be addressed somewhat by observing that the difficulty of implementing $U_R$ is actually closely related to another complexity class known as Quantum Statistical Zero-Knowledge (\textcs{QSZK}). The idea of a zero-knowledge proof is best explained by example. 

Consider the problem of determining whether two graphs $G_1 = (V_1,E_1)$ and $G_2 = (V_2,E_2)$ are isomorphic, that is, whether there exists a permutation $\pi$ of the vertices of $G_2$ turning $G_2$ into $G_1$. There is currently no polynomial time classical or quantum algorithm known for the graph isomorphism problem.
Suppose, however, that some inventive computer scientist \emph{claimed} to be able to solve the graph isomorphism but jealously guarded his secret abilities. Would he be able to convince you that two $G_1$ and $G_2$ are isomorphic without revealing any information about the isomorphism? Yes! The computer scientist begins by randomly permuting the vertices of $G_1$, sending you the resulting graph $G_3$. At that point, you flip a coin and, depending on the outcome, challenge him to exhibit an isomorphism to either $G_1$ or $G_2$. He will be able to succeed if the graphs really were isomorphic but will necessarily fail half the time otherwise. After a few repetitions of the process, you will become convinced of the existence of the isomorphism between $G_1$ and $G_2$ without learning anything at all about its structure.

There are different ways to formalize the idea of a zero-knowledge proof, leading to potentially different complexity classes. The version relevant here is known as the class Statistical Zero-Knowledge (\textcs{SZK})~\cite{goldwasser1989knowledge}. In the quantum mechanical analogue, the two participants, usually known as the prover and the verifier, would exchange qubits rather than bits, with the resulting class called \textcs{QSZK}~\cite{watrous2002limits}. There is absolutely no constraint on the computational power of the prover, but the verifier can only perform polynomial time quantum computations. Moreover, only a statistically negligible amount of information should leak from the prover to the verifier. \textcs{QSZK} is the set of computational problems with yes/no answers for which such a prover can always convince the verifier of yes instances but will fail with high probability for no instances. It is known that the quantum model is at least as powerful as the classical one: $\textcs{SZK} \subseteq \textcs{QSZK}$~\cite{watrous2009zero}.
\textcs{QSZK} also trivially contains \textcs{BQP}, the class of problems that can be solved on a quantum computer: for such problems, the verification can be done directly using the computer itself without any need for a clever discussions with a prover. \textcs{QSZK} should therefore be understood as the set of problems whose yes instances can be reliably identified using a quantum computer with the help of an all-powerful prover, albeit one who is both secretive and dishonest. To assert that all the problems in \textcs{QSZK} can be solved in quantum polynomial time, that \textcs{QSZK} = \textcs{BQP}, is to assert that the prover is ultimately no help at all.

Given some arbitrary polynomial-sized quantum circuit $U_{dyn}$ acting on three systems $B$, $H$ and $R$ such that $|\psi\rangle_{BHR} = U_{dyn} | 0 0 0 \rangle_{BHR}$ with $|\psi\rangle_{BHR}$ maximally entangled between $B$ and $HR$,
determining whether maximal entanglement with $B$ can be decoded from $R$ is a well-defined computational problem. Call it the \textcs{Error Correctability} problem.\footnote{The channel constructed in the previous subsection shows how to relate this definition to more conventional error correction.} Note that \textcs{Error Correctability} has a quantum statistical zero-knowledge proof, which simply consists of having the prover implement the quantum error correction operation on $R$ (not caring that it might take exponential time) and having the verifier check the result. In fact, the problem of determining whether noise is correctable in this sense is \emph{complete} for \textcs{QSZK}, meaning that any other problem in \textcs{QSZK} can be efficiently mapped onto a version of the error correction problem~\cite{hayden2013qszk} (see also section 7 of \cite{hayden2012two}). 

Suppose now that, given a circuit for an arbitrary correctable $U_{dyn}$,  it were possible to efficiently find and implement the error correction procedure. In that case, the entire zero-knowledge proof for \textcs{Error Correctability} described above could be implemented efficiently on a quantum computer. In the case of yes instances, the procedure would produce verifiable maximal entanglement with $B$. In the case of no instances, no such entanglement could be produced regardless of the decoding procedure attempted. Moreover, since Error Correctability is \textcs{QSZK}-complete, that means that every problem in \textcs{QSZK} could be solved on a quantum computer: being able to efficiently decode noise whenever it is correctable would imply that \textcs{QSZK} = \textcs{BQP}. (It is important to remember, however, that simply determining whether some errors are correctable could be much \emph{easier} than actually correcting them.)

Crucially, \textcs{Error Correctability} remains \textcs{QSZK}-complete even if $B$ consists of only a single qubit, unlike the \NP-hardness result for stabilizer decoding discussed earlier.  The question at the core of this article is whether the decoding can be performed in time polynomial in the size of the circuit for $U_{dyn}$.  We have seen that being able to do so would imply that \textcs{QSZK} = \textcs{BQP} if $U_{dyn}$ represented arbitrary correctable noise.

Since the real $U_{dyn}$ describing black hole evaporation is very special, we could ask whether its known properties are so unusual as to undermine the argument. Specifically, for sufficiently late times, we expect maximal entanglement between $BH$ and $R$. It is possible to a certain extent to achieve the same thing using arbitrary $|\psi\rangle_{BHR} = U_{noise} | 000 \rangle_{BHR}$ by simply working with $k$ copies of $|\psi\rangle$. The resulting state $|\psi\rangle^{\otimes k}$ rapidly converges to one with near-maximal entropy concentrated in the ``typical subspace'' of  $B^{\otimes k}H^{\otimes k}$~\cite{schumacher1995quantum}. This property is similar to true maximal entanglement generated by $U_{dyn}$, albeit slightly weaker. 

The conviction that $\textcs{P} \neq \NP$ has developed over several decades of research in algorithm design and complexity theory. The belief that \textcs{QSZK}-complete problems cannot be solved efficiently on a quantum computer is admittedly less well-founded but does have some algorithmic and complexity theoretic support. 

Researchers have been working for twenty years on the design of efficient quantum algorithms and some problems have stubbornly resisted attack. In particular, Shor's factoring algorithm naturally extends to an efficient quantum algorithm for the more general \textcs{Abelian Hidden Subgroup} problem~\cite{kitaev1995quantum}. Researchers have been trying consistently since then to attack the non-Abelian version of the problem but with only very limited success~\cite{bacon2005optimal,kuperberg2005subexponential,ivanyos2008efficient}. (Note that the non-Abelian version includes the graph isomorphism problem discussed above as a special case~\cite{ettinger1999quantum}.) Large classes of strategies based on Shor's Fourier-sampling approach are known to fail~\cite{moore2008symmetric,moore2010impossibility}. 

Given all the fruitless effort that has gone into trying to find an efficient quantum algorithm for solving the \textcs{non-Abelian Hidden Subgroup} problem, researchers have begun to suspect that no such algorithm exists. Moore, Russell and Vazirani (MRV) took one step further and defined a classical invertible function that is efficient to evaluate but hard to invert on a quantum computer under the assumption that there is no efficient quantum algorithm for \textcs{non-Abelian Hidden Subgroup}~\cite{moore2007classical}. Their construction is easily adapted to rule out efficient quantum algorithms for decoding efficiently encoded quantum error correcting codes under the same assumption.
Structurally, the function is parametrized by a list of $m$ vectors $V$ over $\mathbb{F}_q^n$. The MRV function $f_V$ takes a matrix $M \in \text{GL}_n(\mathbb{F}_q)$ to $MV$, with the output returned as an unordered list. ($m$ is selected to be only slightly larger than $n$, which is sufficient to ensure that the function is injective with high probability.) 

Instead of returning an unordered list, however, the output vectors could equivalently be ordered but permuted by an unknown permutation $\pi$, which can be used to define the following isometry:
\begin{equation}
U : |{M}\rangle_A \mapsto \frac{1}{\sqrt{m!}} \sum_{\pi \in S_m} |\pi\rangle_H |\pi(MV)\rangle_R.
\end{equation}
Then
$(I_B \otimes U)$ acting on a state maximally entangled between $A$ and $B$ efficiently generates a state maximally mixed on $BH$ with the property that the purification of $B$ can be recovered by a unitary acting on $R$ alone, precisely mimicking the key properties of $\sum_{b,h} |b\rangle_B |h\rangle_H U_R |bh0\rangle_R$. If $f_V$ is hard to invert on a quantum computer, however, the decoding unitary can't be implemented in polynomial time. Under the assumption that there is no efficient algorithm for \textcs{non-Abelian hidden subgroup}, however, $f_V$ \emph{is} hard to invert, even for $V$ chosen uniformly at random.

At the level of complexity theory, there is some evidence that \textcs{QSZK}-complete problems cannot be solved using small quantum circuits. (An efficient quantum algorithm corresponds to a small circuit that can itself be laid out efficiently, a further requirement that should arguably be relaxed in discussions of the AMPS paradox.) It is known that determining whether a function is 1-to-1 or 2-to-1 requires exponentially many calls to the function, even for a quantum computer~\cite{aaronson2002quantum}. It is easy to construct a statistical zero-knowledge protocol for the problem, however, in the setting in which the prover knows the function and the verifier is making queries to try and distinguish the 1-to-1 and 2-to-1 cases. Since $\textcs{SZK} \subseteq \textcs{QSZK}$, the protocol lifts to a quantum statistical zero-knowledge proof as well. In this model, therefore, an exponentially large number of queries is required to solve the problem using a quantum computer even though it has a zero-knowledge protocol. Any demonstration that \textcs{QSZK} has small circuits would somehow have to be reconciled with that fact, which essentially rules out any strategy which just directly transforms a zero-knowledge protocol directly into a small circuit. Instead, the demonstration would need to make essential use of some subtle internal structure of the problems contained in \textcs{QSZK}.\footnote{We thank John Watrous for suggesting this argument.}

The conclusion of this section is that by recasting Alice's problem as quantum error correction, we have set it into a framework where there are general arguments that such problems likely take exponential time to solve.  Moreover the practical difficulties of doing the experiment, in particular the problems associated with measuring gravitons, further increase the difficulty of this computational task.  We did not quite manage to prove that her task is {\NP}-hard at fixed $k$, but it is almost certainly at least \textcs{QSZK}-hard and there are strong reasons to believe that such problems can't be solved in polynomial time on a quantum computer.
From the computer science point of view, it would be extremely surprising if implementing $U_R^\dagger$ did not require exponential time.

\section{More General Black Holes}\label{adssect}
For a Schwarzschild black hole the evaporation time scales like the entropy to the $3/2$ power, which is clearly much too fast for Alice to complete a computational task that requires time that is exponential in the entropy.  In this section we consider the AMPS experiment for some more general classes of black holes.

\subsection{Schwarzschild in a Box}
In order for Alice to have any chance of doing the AMPS experiment given our claim of exponential decoding time, she will clearly need some way of slowing down the evaporation of the black hole.  The simplest thing she could imagine is letting the black hole become old and then putting it inside of some sort of reflecting box to prevent it from evaporating.  

Closed finite entropy systems however behave very strangely over times of order $e^S$.  For example over that kind of timescale a gas of particles in a room sometimes finds itself collected up in the corner of the room, and other times find itself spontaneously assembling into a puddle of liquid on the floor.  This is the phenomenon of \textit{Poincar\'e recurrence} \cite{sussrec}.  Indeed $e^S$ is sometimes called the ``classical recurrence time'', to be contrasted with the ``quantum recurrence time'' $e^{e^S}$ which we encountered in section \ref{bigcompsect}.  This nomenclature is a little misleading; after all, quantum mechanics is the reason that the entropy of the gas is finite in the first place, but what it really means is the following: $e^S$ is the time scale over which some finite entropy quantum mechanical system will be ``classically close'' to an order one fraction of some orthogonal basis for its Hilbert space, up to conservation laws.  In other words the probability that repeated measurements in that basis over a time of order $e^S$ at some point give any particular result allowed by conservation laws is order one.  The significantly longer quantum recurrence time, by contrast, is the time it takes for the system to get close in the trace norm to any particular quantum state.  The basic distinction here is that the number of elements in a basis of the Hilbert space is $e^S$, while the number of elements in an $\epsilon$-net of the type discussed in section \ref{bigcompsect} is $e^{e^S}$.  

This means that it will be very difficult to confine the black hole in a box for such a long period of time.  For example there is an order one probability that the black hole will produce a gigantic nuclear warhead and fire it at the side of the box.  Or another black hole.  The black hole will also itself crash into the box every now and then.  Making the box big to try to avoid such things is not allowed because then the black hole will just evaporate into a diffuse gas of particles inside the box and there will be no black hole left when Alice finishes her computation and opens the box.  

Of course every string theorist knows how to avoid the problems with putting a black hole in a box in Minkowski space: we just put it in Anti de Sitter space!  As long as the black hole is large enough the reflecting boundary of AdS space will feed its own radiation back into it fast enough to prevent it from evaporating.  Of course to make the AMPS argument at all we need the black hole to become maximally entangled with some external system, so following a suggestion of Don Marolf we imagine this is done by mining the black hole down to less than half of its initial entropy.\footnote{It is unclear to what extent this argument can be applied to the eternal two-sided AdS black hole, since the thermal bath will make it difficult to extract energy.  Indeed of all black holes this seems to be the least likely to have a firewall.  Its gravity dual is two CFT's which are entangled in just the way that seems necessary to produce a smooth horizon \cite{Maldacena:2001kr}.  When we discuss AdS black holes we will always be imagining one-sided black holes made from some sort of collapse.  Interesting previous work on the interior of AdS black holes includes \cite{Fidkowski:2003nf,Horowitz:2009wm}; it would be illuminating to understand how to ask the firewall question in either of these frameworks.}  But this setup brings with it a new problem; we now need to put Alice, her mining equipment, and her computer in the box as well!  Since the calculation still takes of order the recurrence time, this means that both Alice and her assorted paraphernalia now need to be resistant to nuclear warheads/mini black holes/etc.  Alice could try to avoid this by staying very far away from the black hole, namely exponentially near the boundary, so that from her point of view the recurrences become effectively low energy enough not to affect her.  In doing so however she will be fighting against an effective potential pulling her back to the center.  To do this for enough time to accomplish the computation would require exponentially large amounts of rocket fuel, and the exhaust from burning all that fuel would fall back into the black hole anyway and pollute her experiment.  Trying to use angular momentum to stabilize her orbit would not work because over such long time scales her orbit would rapidly decay via gravitational radiation.  

In fact there is a way to avoid these problems as well, which was suggested to us by Juan Maldacena.  It involves putting a big AdS black hole in a throat geometry where the near horizon region is asymptotically AdS but there is also an asymptotically Minkowski region.  In this way the geometry provides a box with the benefits of both the Minkowski box and the AdS box without the problems of either.  There is still a problem with this setup, but it is more subtle and we return to it later in this section.
\subsection{Reissner-Nordstrom}
Another way Alice could try to extend the lifetime of her black hole is by giving it some charge.  The Reissner-Nordstrom black hole of mass $M$ and charge $Q$ has metric
\be
ds^2=-dt^2f(r)+\frac{dr^2}{f(r)}+r^2 d\Omega_2^2,
\ee
where
\be
f(r)=1-\frac{2GM}{r}+\frac{GQ^2}{r^2}\equiv \frac{(r-r_+)(r-r_-)}{r^2}
\ee
and
\be
r_\pm=GM\pm\sqrt{G(GM^2-Q^2)}.
\ee
Its entropy is 
\be
S=\frac{\pi r_+^2}{G}\equiv \frac{\pi r_+^2}{\ell_p^2}
\ee
and its temperature is
\be
T=\frac{r_+-r_-}{4\pi r_+^2}.
\ee
We see when $M\ell_P=Q$, the black hole is extremal and the temperature is zero.  If we start the black hole with mass above extremality, it will radiate and gradually approach extremality.  The relevant point here is that as it does this, the temperature decreases and semiclassically it appears that the decay takes an infinite amount of time.  More precisely if we define the energy above extremality to be
\be
E=M-Q\ell_p^{-1},
\ee
the energy-temperature relation is
\be\label{exttemp}
E=2\pi^2 Q^3 T^2 \ell_p
\ee
and a simple calculation tells us that, for a black hole which starts with energy $E_0\leq Q\ell_p^{-1}$, we have
\be
E(t)=\frac{E_0}{1+\alpha tE_0Q^{-4}}.
\ee
Here $\alpha$ is some order one numerical constant.  Thus it appears that the black hole takes an infinite amount of time to reach extremality.  There is a well-known problem with this argument however \cite{Preskill:1991tb,maldstrom}, which is that from \eqref{exttemp} we see that when the temperature reaches
\be
T\sim \frac{1}{Q^3 \ell_p}
\ee
the remaining energy $E$ above extremality is no longer bigger than the temperature $T$.  At this point the semiclassical description of the evaporation process breaks down, and quantum gravity is necessary to understand what happens next.  This is related to an instability of $AdS_2$ called fragmentation \cite{maldstrom}, and in the cases where it can be understood in string theory it is always true that the geometry breaks apart into something that has little resemblance to the original Reissner-Nordstrom geometry.

It is easy to see that no matter whether we start with $E$ near extremality or much larger, this instability always sets in well before the exponential of the initial entropy.  For example,  say that we start with $E_0\leq Q$ (setting $\ell_p=1$).  The initial entropy is of order $Q^2$, while the time to reach the instability is of order $Q^7$.  Alternatively if we start with $E_0\gg Q$ then we are back to the Schwarzschild situation where evaporating back down to $E\approx Q \ell_P^{-1}$ takes a time of order $E_0^3\approx S_0^{3/2}$, and the additional time to get down to the instability is $Q^7\ll S_0^{7/2}$.  The total evaporation time is always bounded by a polynomial in $S_0$.  Thus charged black holes, near extremal or otherwise, are of no use to Alice.

\subsection{Near Extremal AdS Throat}
Recall the $AdS_d$ Schwarzschild geometry
\be\label{AdSsch}
ds^2=-f(r)dt^2+\frac{dr^2}{f(r)}+r^2d\Omega_{d-2}^2,
\ee
with
\be
f(r)=1+\left(\frac{r}{R}\right)^2-\frac{\alpha}{r^{d-3}}.
\ee
A scalar field on this background feels an effective potential.  In particular if we look for solutions of the Klein-Gordon equation of the form
\be
\phi\equiv r^{-\frac{d-2}{2}}e^{-i\omega t}Y_{\ell}(\Omega_{d-2})\Psi_{\omega \ell}(r),
\ee
it is not hard to see that $\Psi_{\omega \ell}$ must obey a Schrodinger-type equation
\be
-\frac{d}{dr_*^2}\Psi_{\omega \ell}+V_{eff}(r)\Psi_{\omega\ell}=\omega^2 \Psi_{\omega \ell}.
\ee
Here $r_*$ is a ``tortoise'' type coordinate obeying
\be
r_*'(r)=f^{-1},
\ee
and the effective potential is
\begin{align}\nonumber
V_{eff}(r)=R^{-2}\bigg(1+\left(\frac{R}{r}\right)^2&-\frac{\alpha R^2}{r^{d-1}}\bigg)\bigg[\left(m^2+\frac{d(d-2)}{4R^2}\right)r^2\\
&+\left(\ell(\ell+d-3)+\frac{(d-2)(d-4)}{4}\right)+\frac{(d-2)^2}{4}\cdot \frac{\alpha}{r^{d-3}}\bigg].\label{AdSPot}
\end{align}
The details of this potential do not matter, but we see that it vanishes linearly at the only real positive root of $f(r)$, that is at the horizon, and that it grows quadratically with $r$ at large $r$. Near the horizon the solution then behaves as $\Psi_{\omega\ell}\sim e^{i\omega(\pm r_*-t)}$, while at large $r$ we have $\phi\sim r^{-\Delta_{\pm}}$ with the usual AdS/CFT formula
\be
\Delta_{\pm}=\frac{d-1}{2}\pm\frac{1}{2}\sqrt{(d-1)^2+4R^2m^2}.
\ee

The idea of this section is to cut off this geometry at some large value of $r$ and sew it onto Minkowski space, after which the effective potential \eqref{AdSPot} would go back to zero provided we set the scalar field mass $m^2$ to zero.  The black hole would then be able to decay via massless quanta tunneling out of this potential into the Minkowski region, and by choosing the crossover value of $r$ to be large we could adjust the decay time independently of the entropy of the black hole.  We can also ``outsource'' the computation by putting the computer out in the Minkowski region, which allows us to buy a large redshift factor enhancement in the time it takes to do the computation from the point of view of Alice living down by the black hole.  

There is a new problem with this construction however, which is that any attempt to send the result of the computation from the Minkowski region back down the throat to the vicinity of the black hole has to get back through the potential barrier.  The signal the computer sends down the throat will need to be very low energy, so its absorption probability will be exponentially small.  Nonetheless one could imagine trying to send the signal repeatedly, hoping that eventually one of the times it will get through.  This approach gets much harder as the size of the signal we wish to send increases, but unlike the previous examples it is not obvious that it cannot be done and we need to analyze it more carefully.  

\subsubsection{The Brane Setup}
To quantitatively test the feasibility of using a throat to evade the computational hardness of decoding, we need a specific example of this type of geometry.  In string theory there is a standard way of producing throat geometries with the desired properties by stacking branes; for example D3 branes in type IIB string theory or M5 branes in M-theory.  To realize the geometry \eqref{AdSsch} explicitly we would need a configuration of branes which is spherically symmetric and stable.  Spherical symmetry is a bit inconvenient because branes in spherical configurations tend to collapse under their own tension unless there is something else supporting them.  A simple thing we could do is start with the $AdS_3\times S^3\times T^4$ solution of IIB supergravity with RR flux, itself the near horizon limit of the $D1-D5$ system, and then wrap some $D3$ branes on the $S^3$.  This throat would not be asymptotically Minkowski, but we could easily arrange for the curvature radius of the $AdS_3$ to be much larger than the curvature radius of the $AdS_5$ near the $D3$ branes.  Rather than try to make this construction work in detail, we will instead consider a simpler setup in which the black hole has planar symmetry instead of spherical symmetry.  

One of the best-known solutions of ten-dimensional IIB supergravity is the extremal planar black 3-brane \cite{Horowitz:1991cd}, with metric
\be\label{extbb}
ds^2=Z(r)^{-1/2}\left(-dt^2+d\vec{x}^2\right)+Z(r)^{1/2}\left(dr^2+r^2 d\Omega_5^2\right).
\ee
Here
\be
Z(r)=1+\left(\frac{R}{r}\right)^4,
\ee
and $R$ is a parameter of the solution.  The string theory interpretation of this solution is that it gives the backreacted geometry in the presence of $N$ D3 branes, with
\be
R^4=4\pi g N \ell_s^4\sim N \ell_p^4
\ee
where $g$ is the string coupling, $\ell_s$ is the string length, and $\ell_p$ is the ten-dimensional Planck length.  By looking at $Z(r)$ we see that this geometry indeed has the property that for $r\ll R$ it behaves like $AdS_5$ in Poincar\'e coordinates, times an extra $S^5$ of constant radius, while for $r\gg R$ it becomes ten dimensional Minkowski space.

To get something like a black hole down the throat we need to put this solution at finite temperature, and to get the entropy to be finite we need to compactify the spatial $\vec{x}$ directions.  Although it will not be explicit in our equations, we will choose anti-periodic boundary conditions for the fermions around the compact dimensions.\footnote{The reason for this choice is that compactifying with periodic boundary conditions preserves supersymmetry, which introduces an instability of the throat.  Supersymmetry ensures there is no potential energy cost for separating the D3 branes that make up the throat.  In CFT language the dual gauge theory is unable to pick a vacuum and its zero modes wander freely on its moduli space.  Antiperiodic boundary conditions break supersymmetry and generate a potential that keeps the branes together.  We thank Igor Klebanov and Juan Maldacena for discussions of this point.}  To add some temperature we just need to consider the non-extremal version of the solution \cite{Gubser:1996de}:
\be\label{nonext}
ds^2=Z(r)^{-1/2}\left(-f(r)dt^2+d\vec{x}^2\right)+Z(r)^{1/2}\left(\frac{dr^2}{f(r)}+r^2 d\Omega_5^2\right),
\ee
where $Z(r)$ is now 
\be
Z(r)=1+\lambda \left(\frac{R}{r}\right)^4,
\ee
with 
\be
\lambda=\sqrt{1+\frac{1}{4}\left(\frac{r_0}{R}\right)^8}-\frac{1}{2}\left(\frac{r_0}{R}\right)^4
\ee
and
\be
f(r)=1-\left(\frac{r_0}{r}\right)^4.
\ee

To put the black hole far down the throat we clearly want $\frac{r_0}{R}\ll1$, and in this limit the entropy of the black hole is
\be\label{ent}
S=\frac{\Omega_5 L^3r_0^3R^2}{4G}\sim \frac{L^3 r_0^3 R^2}{\ell_p^8}
\ee
and the temperature is
\be\label{temp}
T=\frac{1}{\pi}\frac{r_0}{R^2}.
\ee
Here $\Omega_5$ is the volume of a unit $\mathbb{S}^5$ and $L$ is the periodicity of the $\vec{x}$ directions.  In the same limit the ADM energy \footnote{For an asymptotically flat geometry, in coordinates where the metric is $\eta_{\mu\nu}+h_{\mu\nu}$ with $h_{\mu\nu}$ small, the ADM energy is defined \cite{PhysRev.122.997} as $\lim_{r\to \infty}\frac{1}{16\pi G}\int dA n^{i}\left(\partial_j h_{ij}-\partial_i h^j_{\phantom{j}j}\right)$.  Here the integral is over the $\mathbb{S}^5\times \mathbb{T}_3$ at infinity.} is
\be
E_{ADM}=\frac{\Omega_5L^3}{4\pi G}\left(R^4+\frac{3}{2}r_0^4\right).
\ee
One can check that when $r_0\to 0$ this reproduces the correct D3-brane tension.  The energy above extremality is
\be\label{eng}
E=\frac{3\Omega_5 L^3 r_0^4}{8 \pi G}\sim \frac{L^3 r_0^4}{\ell_p^8}.
\ee
It is straightforward to check that the instability encountered in Reissner-Nordstrom does not happen here; in fact we have
\be\label{engent}
E=\frac{3}{2} S T,
\ee
so the energy will be bigger than the temperature until the black hole has Planckian area.  We may then worry that the black hole evaporation time is independent of the intial entropy, as we worried with Reissner-Nordstrom, but there is now a new phenomenon which comes to the rescue.

\subsubsection{Hawking-Page Transition for Toroidal Black Holes}
For spherical black holes in AdS with metric \eqref{AdSsch}, it is well known \cite{Hawking:1982dh,Witten:1998zw} that for small enough values of $\alpha$ the black hole is unstable to decay by Hawking radiation.  The crossover point is when the temperature is of order  $R^{-1}$.\footnote{More carefully this is the temperature below which the black hole no longer dominates the thermal ensemble.  In the microcanonical ensemble the energy below which a single black hole is actually unstable is lower by some power of $R/\ell_p$, but this distinction will not be important for us for reasons we explain below.  We thank Juan Maldacena for explaining this distinction to us.}  This phenomenon provides a natural endpoint for the type of decay we discussed in the previous section; the black hole will very slowly radiate energy up the throat and into Minkowski space until it reaches the critical temperature, after which it will decay essentially immediately into low energy quanta in the AdS-region of the throat.  Does something similar occur for our toroidal black hole as well?  The answer is yes, but as we now describe the analysis has a few details that differ from the spherical case.  This transition has been previously discussed by \cite{Nishioka:2009zj}.

Far down the throat, our geometry becomes a compactified version of a general solution called a black AdS brane.  In $AdS_d$ the metric for this solution is\footnote{In this section we set the AdS radius $R$ to one.}
\be\label{bb}
ds^2=r^2\left(-f(r)dt^2+d\vec{x}^2\right)+\frac{dr^2}{r^2f(r)},
\ee
with 
\be
f(r)=1-\frac{\alpha}{ r^{d-1}}.
\ee
In the spherical case the action of the Euclidean version of the solution \eqref{AdSsch} can be compared to the action of another Euclidean solution which has the form \eqref{AdSsch} again but with $\alpha=0$.  In both cases the Euclidean time $\tau=it$ is compactified.  The competition between these two solutions is the source of the Hawking-Page transition.  In the toroidal setting there is an analogous construction of a second solution, where we set $\alpha=0$, but it turns out always be subdominant to the black brane solution  \eqref{bb}.  There is another solution however with the same boundary conditions; we can take the ``emblackening factor'' $f(r)$ and move it from in front of $-dt^2$ to one of the planar coordinates:\footnote{Had we chosen supersymmetric boundary conditions in the spatial directions this solution would not exist.}
\be\label{adssol}
ds^2=r^2\left(-dt^2+f(r)dx^2+(d\vec{x}^\perp)^2\right)+\frac{dr^2}{r^2f(r)}.
\ee
This geometry is sometimes called the AdS Soliton.  It has no region behind the ``horizon'' at $r=r_0$, instead it caps off smoothly provided that we choose the correct value of $r_0$ as a function of the periodicity $L$.  When we continue to Euclidean time we can set the $\tau$ periodicity of the AdS soliton \eqref{adssol} freely, but for the black brane \eqref{bb} we must choose $r_0$ to be consistent with the $\tau$ periodicity for the Euclidean geometry to be smooth.  An important subtlety is that for a given periodicity of the circle at the boundary, the correct value of $r_0$ and also the coordinate periodicity of $\tau$ are different in the two solutions.\footnote{Actually in Euclidean signature they are really just the same solution with different parameters, so the calculation only needs to be done once.}  The Euclidean action is
\be
S=-\frac{1}{16\pi G}\left(\int d^{d}x \sqrt{-g}\left(R+(d-1)(d-2)\right)+2\int d^{d-1}x \sqrt{\gamma} K\right),
\ee
where $\gamma$ is the determinant of the induced metric at the boundary and $K$ is the trace of the extrinsic curvature.  Making the action finite involves cutting off the geometry at some large $r=r_c$, and then carefully matching the boundary geometry on the regulator surface in the two cases.  We will not present the details explicitly here since they are fairly standard in the literature. (See \cite{Witten:1998zw,Anninos:2012ft} for examples.) The result is that the finite parts of the actions are
\be
-S_{BB}=\frac{1}{16\pi G}\left(\frac{4\pi}{d-1}\right)^{d-1} (LT)^{d-2}
\ee
for the black brane and 
\be
-S_{sol}=\frac{1}{16\pi G}\left(\frac{4\pi}{d-1}\right)^{d-1} \left(LT\right)^{-1}
\ee
for the AdS soliton.  Thus at high temperatures compared to $L^{-1}$ the black brane wins while at low temperatures the AdS soliton wins.  This then is the effect that we want; as the black hole radiates it will eventually undergo a transition to some other type of geometry with no horizon and Alice will no longer be able to test AMPS.  The dual field theory interpretation of this is clear; it is the same large $N$ phase transition as in the spherical case, just studied with different spatial topology.\footnote{As in the spherical case, in the microcanonical ensemble the energy at which the black hole is actually unstable is somewhat lower than this.  This decrease depends only on the parameters of the AdS region of the geometry however, and is insensitive to the total length of the throat.  Since we will need the total length of the throat to exponential in the entropy in the following section, which will lead to an exponentially long decay time, the additional time to get from $T\sim L^{-1}$ to the actual instability will be negligible compared to the time to get down to $T\sim L^{-1}$ in the first place.} 

\subsubsection{Time Scales}
In this section we work out the time scales for sending signals down the throat from the Minkowski region to the vicinity of the black hole horizon at $r_0$ in \eqref{nonext}, as well as the time to evaporate down to the transition temperature just discussed in the previous section.  Since we will ultimately compare these time scales to the computation and recurrence times, both of which are exponential in the entropy of the black hole, we will be focused on extracting only the pieces of them which are exponential in entropy.  Specifically we will write $S_0$ to mean the entropy of the black hole just after the Page time, so that it is also roughly the size of the radiation and one half of the entropy of the original black hole.  If we had started the computer any later it would just have made the task more difficult, and we want to give Alice a fair shot.  

We will see shortly that for the decay to be slow enough for Alice to have a chance at computing, we will need the temperature to be exponentially small in the entropy, perhaps with some coefficient in front of $S_0$ in the exponent.  From equation \eqref{temp} this means we will need $r_0$ to be exponentially small.  From equation \eqref{ent} we see that to keep the entropy fixed in the same limit we will need $L$ to be exponentially large such that $Lr_0$ is fixed. Looking at $\eqref{eng}$ we see that the energy above extremality will then be exponentially small througout the decay process.  Having fixed $S_0$ we can also derive an interesting bound on the AdS radius $R$ in Planck units: we know that we must have $T>L^{-1}$ to avoid starting the computation below the phase transition discussed in the previous section, so we must have
\be
(TL)^3=\left(\frac{r_0L}{R}\right)^3=S_0\left(\frac{R}{\ell_p}\right)^{-8}>1.
\ee
Thus the AdS radius in Planck units is bounded by a polynomial in $S_0$ and we can neglect it in most equations. 

To understand how hard it is to send some particular quanta down the throat, we need to compute its absorption probability $P_{abs}$.  This probability will be a function of the frequency $\omega$ of the quanta of interest, and in the limit $\omega R\ll1$ it can often be computed analytically \cite{Das:1996we}.  For massless scalars in the extremal black three brane \eqref{extbb} this problem was studied by Klebanov in \cite{igor1}.\footnote{Computing these absorption factors in the extremal background is an excellent approximation for our purposes since $r_0\ll R$ and the potential is very close to extremal.}  A simple generalization of his result shows that for quanta with frequency $\omega$, angular quantum number $\ell$ on the $\mathbb{S}^5$, and momentum $k$ in the planar direction, the absorption probability is
\be\label{Pabs}
P_{abs}\sim \left(\sqrt{\omega^2-k^2}R\right)^{8+4\ell}.
\ee
In sending signals down the throat, we need to use low enough energy to avoid our signals backreacting significantly on the throat.  Certainly a necessary condition is that we need $\omega<E$, and looking at \eqref{engent} this means we need $\omega\sim T$ up to a power of the entropy, which as usual we ignore.  This means that we must use quanta of exponentially low energy to send any messages; the absorption probability \eqref{Pabs} will thus be exponentially small.  It is still possible to send a message, but we must try many times.  Since it takes an energy $\omega^{-1}$ just to produce a message of energy $\omega$, to have any chance of success sending the message we need a time of order
\be
t_{msg}\sim \frac{1}{\omega P_{abs}}.
\ee
Clearly we have the best chance of sending a message using scalars if we set $k=\ell=0$.  Other types of communication will have different absorption probabilities. For example, we show in appendix \ref{stringapp} that the absorption probability for sending messages down a string threading the throat by moving the string along the $\mathcal{S}^5$ is proportional to $(\omega R)^2$.  Apparently this is a better method of communication than the massless scalar, although we will see it is still not good enough to be of use to Alice.\footnote{We thank Joe Polchinski for suggesting a stringy telephone.}  In general we will parametrize low-energy absorption probabilities as
\be\label{Pgen}
P_{abs}=(\omega R)^b,
\ee  
so neglecting all factors polynomial in $S_0$ we can estimate
\be\label{tmsg}
t_{msg}\sim \frac{1}{T^{b+1}}.
\ee
The units here are provided either by powers of $\ell_p$ or $R$, it doesn't much matter.  To be concrete we can evaluate the temperature at the same time that we defined $S_0$, which was just after the Page time.

Absorption probabilities are also important in understanding how long it takes for the black hole to evaporate.  As Hawking showed in his original paper \cite{Hawking:1974sw} the energy flux out of a black hole is
\be\label{evaprate}
\frac{dE}{dt}=-\sum_n\int \frac{d\omega}{2\pi}\frac{\omega P_{abs}(\omega,n)}{e^{\beta\omega}-1},
\ee
where the sum on $n$ is over different modes.  It is often the case that only a particular mode contributes significantly, for example for Schwarzschild black holes it is only the $\ell=0$ mode.  The low energy absorption cross section for Schwarzschild is proportional to $(2GM\omega)^2$, from which one can use \eqref{evaprate} to motivate the usual ``Stefan-Boltzmann'' assumption for the decay rate.  In that case the low energy approximation breaks down before the peak of the integrand and numerical analysis is necessary to compute the prefactor correctly \cite{pageevap}, but for us the temperature is very low compared to $R^{-1}$ so using the low energy approximation for $P_{abs}$ throughout is justified.  The intuition of \eqref{evaprate} is quite simple; the thermal factor is just the expected occupation number of the near horizon modes, and by a basic fact about one-dimensional scattering theory the probability of absorption in from the outside is the same as the probability of transmission out from the inside.  

For the absorption probability \eqref{Pabs} the decay is dominated by $\ell=0$ modes, but it is necessary to include modes of low but finite $k$.  Roughly $k$ is quantized in units of $1/L$, and since we are interested in the region where $TL\gg1$ we will have $\omega\gg L^{-1}$.  We can then convert the sum over discrete modes into an integral over $k$ and write
\be
\frac{dE}{dt}=-\int \frac{\omega d\omega}{2\pi}L^3\int_{|k|<\omega}\frac{ d^3k}{(2\pi)^3}\frac{(\omega^2-k^2)^4 R^8}{e^{\beta\omega}-1}\sim -L^3R^8\int \frac{d\omega}{2\pi}\frac{\omega^{12}} {e^{\beta\omega}-1}\sim -L^3R^8T^{13}.
\ee
More generally we can write
\be\label{genevap}
\frac{dE}{dt}\sim-\int \frac{d\omega}{2\pi}\frac{(L\omega)^a(R\omega)^b\omega}{e^{\beta\omega}-1}\sim -L^a R^b T^{a+b+2},
\ee
where the parameter $b$ is the same as in \eqref{Pgen} and the parameter $a$ accounts the phenomenon just encountered for the scalar.  For the string we discuss in appendix \ref{stringapp} we have $a=0$, $b=2$.  

To find the evaporation time we need to integrate \eqref{genevap} to find the energy as a function of time.  As the decay proceeds $L$ cannot change because it is fixed by the boundary conditions at $r\to \infty$, so it will be $r_0$ that gradually decreases.  We integrate from initial energy $E=S_0 T$ down to final energy $E=\left(\frac{R}{\ell_p}\right)^8 L^{-1}$.  The details of the integral depend on whether  $a+b-2$ is positive, negative, or zero, but the final power of $T$ does not.  Indeed we find
\be\label{tevap}
t_{evap}\sim \frac{1}{T^{b+1}},
\ee 
where again we can make up the dimensions with either $R$ or $\ell_p$ without affecting the exponent in the entropy.  By comparing \eqref{tevap} to \eqref{tmsg} we see that the evaporation time is always the same order in $T$ as is the time to send any signal at all!\footnote{It is the same $b$ that appears in both because whichever $b$ is smallest will control the decay rate and also give the highest probability of success for sending messages.}  Which one is bigger depends on the prefactors we omitted, but there are definitely cases where $t_{msg}<t_{evap}$ so this fact by itself, although certainly troubling, is not enough to kill the experiment.  

In section \ref{compsec} we argued that Alice's quantum computation takes a time of order $e^{2S_0}$.\footnote{The reader should not be confused by us writing $e^{2S_0}$ here and $2^{2n}$ there. Previously $n$ was the entropy in base-2 logarithm while $S_0$ here is the entropy in the natural logarithm.  Also our general result there was $t_{comp}\sim2^{n+k+m}$, but since we are here letting Alice start just after the Page time we have $k+m\approx n$.}  For now we will be a little more general and write this as
\be
t_{comp}\sim e^{\alpha S_0}.
\ee
The recurrence time is basically $e^{S_0}$, but we need to include a red-shift factor to account for the extremely low energy of the states involved in the recurrences.  Thus
\be
t_{rec}\sim T^{-1} e^{S_0}.
\ee

With these estimates we are finally in a position to assess the viability of Alice's experiment.  For the computation to finish before the black hole evaporates we need $t_{comp}<t_{evap}$, which implies
\be
T < e^{-\frac{\alpha}{b+1}S_0}.
\ee
This confirms our earlier claim that the temperature needs to be exponentially small in the entropy.  To be able to send a message down the throat in less than a recurrence time we need $t_{msg}<t_{rec}$, which implies
\be
T> e^{-\frac{1}{b}S_0}.
\ee
The condition that $t_{comp}<t_{rec}$ gives
\be
T<e^{(1-\alpha)S_0}.
\ee
It is straightforward to see that all three of these can be satisfied only if 
\be
b<\frac{1}{\alpha-1}.
\ee
In the text we saw that if $U_R$ is completely general then $\alpha=2$, in which case the experiment can be done only if $b<1$.  Neither the string nor the free scalar field are close to this, and actually there is a simple argument that no scalar field of any kind can satisfy this inequality.  The coefficient $b$ in the absorption factor is related to the conformal dimension of the operator that the scalar couples to in the CFT dual as $b=2\Delta$ \cite{igor2}, so the unitarity bound $\Delta>1$ in four dimensions precludes $b<2$.  A similar argument can perhaps be constructed for the defect operators that couple to the ends of general strings but we have not tried to do so.  

It is interesting that were the computer able to decrease $\alpha$ it would make it easier to satisfy these inequalities.  In fact we don't see any particular reason why improved algorithms shouldn't be able to use special features of the black hole dynamics to decrease $\alpha$ by some order one factor.  If $\alpha$ could be decreased below $3/2$, the string might become an effective method for communicating down the throat.  We stress however that this is not sufficient for doing the AMPS experiment, it is only necessary.  For one thing even if the ``true'' $\alpha$ could be decreased by algorithms, our discussion below equation \eqref{entguess} suggests that, because of coarse-graining, $\alpha$ should be \textit{increased} by some order one factor.  More significantly, being able to send one piece of classical information is not enough to do the strongest version of the AMPS experiment.  That requires us to send a particular quantum state which purifies $B$.  Preserving the coherence of this state would presumably require some sort of apparatus (also made out of fluctuations on the string) which would also have to make it through the barrier, and even without the apparatus we probably want $k$ to be at least a little bit bigger than one to be able to build up any kind of statistics.  Getting all of these things to make it through the barrier at once probably requires us to raise $t_{msg}$ by some order one power, which would help compensate for an $\alpha$ that has been decreased by clever algorithms.  Even this is not enough however; the purification of $B$ is more likely to be partially reflected than to get all the way through, which means that it will be partially reflected many times before most of it gets through.  In fact getting slightly more than half of it through is enough since somebody living down the throat can do error correction to restore the other half, but each time that more than half of it is reflected the person outside will need to do error correction before trying to send it again.  This will usually succeed, but there is a small $S_0$-independent probability it will fail.  Since we need the correction procedure to work every time in order to continue sending the correct state, this means that eventually it will fail.  For these reasons we are quite confident that the ``strong'' AMPS experiment where Alice carries the purification of the black hole in with her can't be done in this setup.

There is a weaker version of the AMPS experiment where the state of $R_B$ is measured out in the Minkowski region at the end of the quantum computation and what is sent down the throat to Alice is just a classical record of the result of that measurement. This would be sufficient, for example, to implement a Bell inequality test using the state entangled between $B$ and $R_B$. Because classical information can be cloned, sending the record of the measurement outcome to Alice is easier than sending $R_B$ itself; she doesn't have to worry about error correction and she can send multiple copies at once. Sending exponentially many copies of the information down the throat at once is dangerous from a back-reaction point of view however.  For example, to do this using strings would require an exponential number of strings, all parallel and located at different values of $x$, which would become extremely dense down the throat; the distances between the strings would become exponentially sub-Planckian.  For massless scalars the total number of modes we can use without backreaction is, at the level of exponential factors,  of order $(LT)^3$.  Since $L \sim T^{-1}$, using all of these modes at once is of no help in trying to beat the exponential in $t_{msg}$.  

Given the possibility of $\alpha<2$ we are not decisively able to rule out this `weak'' AMPS experiment, but at a minimum we are comfortable interpreting this section as casting serious doubt on the feasibility of using an AdS throat to facilitate an AMPS experiment.\footnote{Steve Shenker has suggested that another way to buy a little more ``room'' for dealing with $\alpha<2$ is to notice that the classical recurrence time decreases under coarse-graining.  For example if one considers a system of $N$ particles on a line with $L$ lattice points, $S=N \log L$ and $e^S$ is the time one has to wait to find all $N$ particles on a single lattice point.  To find all of them on some particular set of $\sqrt{L}$ lattice points however we only need to wait a time of order $e^{S/2}$.  Also we have so far essentially treated Alice as a ``probe observer'', but Borun Chowdhury has pointed out that Alice and her equipment will have their own recurrence time which is  probably significantly shorter than that of the black hole.}  

\section{The Structure of the Hilbert Space} \label{options}
The main argument of this paper is now complete, and although the paper is already long, we can't resist making a few comments about the possible implications of our results for how to think about the interior of a black hole.  In the introduction we briefly discussed two alternatives, ``strong complementarity'' and ``standard complementarity'', for how to think about Alice the infalling observer's quantum mechanics.  It is very important to decide which, if either, of these frameworks is the correct way to think about black hole interiors.  In this section we assess the status of each in turn in light of our computational arguments.  This section has substantial overlap with a paper by Susskind \cite{Susskind:2013tg} appearing simultaneously with this one, and which explains some of these ideas in more detail.

The least restrictive idea for how to think about Alice is to imagine that she has her own quantum mechanics, possibly approximate since she encounters a singularity later, which is a priori independent of the quantum mechanics of an observer at infinity like Charlie.  This type of theory has been argued for by Banks and Fischler for a while, who attempt to realize it precisely as quantum mechanics with no approximations for anybody, in a formalism called ``holographic spacetime'' \cite{Banks:2001px,Banks:2011av}.  The basic idea will be discussed using as illustration figure \ref{alice} above.  In this framework it seems necessary for consistency \cite{bousso,harlow} for Alice and Charlie to agree about the results of measurements in the green region of the figure.  It is fairly clear in this setup that if Alice indeed cannot decode the radiation before jumping in, we are free to change her state in a way that produces no observable contradiction with the fact that Charlie, who is able to decode, will later conclude that that $R_B$ was entangled with $B$.  Charlie, not having access to the degrees of freedom behind the horizon, will not be able to check that $B$ is entangled with $A$.  Moreover a previous objection to strong complementarity \cite{Almheiri:2012rt}, that it required some sort of discontinuity in the experiences of a sequence of observers who jump in at different times, is not relevant since the only thing that determines whether or not $R_B$ can be decoded is whether or not the observer ever falls in.  So Alice's inability to decode apparently to allows strong complementarity to be consistent without firewalls.

Although strong complementarity is in some sense straightforward, it is rather unsatisfying.  Each observer having her own description of the universe, approximate or not according to taste, and with no clear precise relationship between them, seems to us like a rather inelegant fundamental framework.  
In particular it is to be contrasted with AdS/CFT \cite{Maldacena:1997re,Witten:1998qj,Gubser:1998bc}, where there is a single Hilbert space and set of operators which is conventionally understood to describe all of the physics in AdS space within a single sharp framework.  It would be reassuring if strong complementarity could be set into such a framework, in which its ambiguities would be understood as arising from measurements that are not precisely well-defined.  Without such an embedding, it would seem like strong complementarity would amount to a step backwards; even in AdS space the CFT would not be a complete description of the physics of an infalling observer.\footnote{The point of view of that follows here is in some respects close to that of \cite{Papadodimas:2012aq}, although we disagree with their assertion that their construction by itself addresses the argument of AMPS.}  As we discussed above this embedding has been called $A=R_B$ in the context of firewalls; in the remainder of this section we sketch a basic proposal for how it could work more precisely.  

\begin{figure}
\begin{center}
\includegraphics[height=5cm]{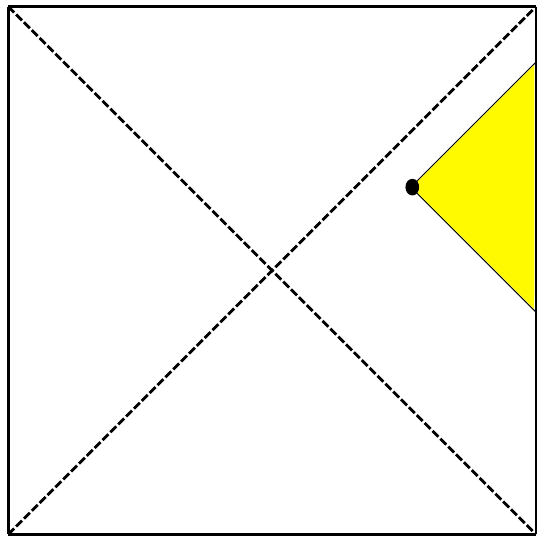}
\end{center}
\caption{The construction of bulk operators by Kabat, Lifschytz, and Lowe for an AdS black hole.  Here the yellow region is a lightcone ending on the operator and extending out to the boundary, and the operator is constructed by integrating a CFT operator over the boundary of the yellow region against a kernel that depends on the position of the operator.  Note that as the operator approaches the horizon the operator becomes sensitive to the entire history on the boundary and thus to the details of the quantum state.}\label{kabat}
\end{figure}

We imagine that there is a single Hilbert space $\mathcal{H}$, which we will think of as the CFT Hilbert space in an AdS setup to be concrete.  To understand Charlie's physics on some spatial slice like the black one in figure \ref{alice}, we need to compute expectation values of some set of operators $C_n$, which approximately commute.\footnote{It is interesting to ask whether approximately commute means up to powers in $N^{-1}$ or up to exponentially small terms in $N$.  We are agnostic about this here.}  Indeed there is a fairly well-known construction \cite{Kabat:2011rz} for constructing these operators in some cases, which we illustrate in figure \ref{kabat}.  

Our proposal for the interior is then that there is another set of operators, which we will call $A_n$'s, which are also mutually commuting with each other and whose expectation values in the same (Heisenberg picture) state used by Charlie describe Alice's experience on the red slice in figure \ref{alice}.  Some of the $A_n$'s are interpreted by Alice as being outside the horizon, and she can also try to construct them using the method of \cite{Kabat:2011rz}.  Consistency then requires that these $A_n$'s are equal to the appropriate $C_n$'s to prevent disagreement between Alice and Charlie about events in the green region of figure \ref{alice}.  Others of the $A_n$'s are interpreted by Alice as being behind the horizon and, as shown in the figure the construction of \cite{Kabat:2011rz}, breaks down in that case.  These $A_n$'s naively do not seem to have low energy interpretations for Charlie.  From the AMPS argument however we know that to have a smooth horizon it must be that there are some operators just outside the horizon, which act on what we've been calling $B$, measurements of which need to be close to perfectly correlated measurements of some of the behind-the-horizon $A_n$'s.  Before the Page time, none of Charlie's $C_n$'s are expected to have this correlation with the $B$ operators, and as recently argued by Susskind, Verlinde, and Verlinde \cite{Susskind:2012uw,Verlinde:2012cy} Charlie can then interpret the $A_n$'s as just being some complicated mess acting on the remaining black hole.  After the Page time, however, Charlie expects the operators acting on $B$ to be perfectly correlated with nonlocal $C_n$'s acting on what we've called $R_B$ in the radiation.  So it must be that from Charlie's point of view the appropriate $A_n$'s now act on the complicated subfactor of the radiation which purifies $B$.  Hence the name $A=R_B$.  This clearly is rather nonlocal, but as shown in the figure the breakdown of the construction of \cite{Kabat:2011rz} suggests that the construction of operators behind the horizon does indeed depend on sensitive details of the state.

This idea is very confusing to interpret however if Alice is able to decode the Hawking radiation, because she then has two low energy observables which she wants to identify with the same quantum mechanical operator on the same Hilbert space.  This is sometimes called cloning, although it isn't really because the theory is quantum mechanical and thus doesn't clone, but it seems like a rather serious problem for the physical interpretation of quantum mechanics.  By doing low-energy manipulations of the Hawking radiation Alice would be able to construct a situation where looking at some localized piece of the Hawking radiation far from the black hole is indistinguishable from looking behind the horizon.  At a minimum this type of observable bizarreness would allow acausal communication, and in any event it doesn't seem particularly less crazy than the idea that there is a firewall.  In the context of the discussion of this paper however, if Alice in principle cannot decode $R_B$ then there does not seem to be any such problem with interpreting $A_n$ as being behind the horizon from Alice's point of view and out in the radiation from Charlie's point of view.  This is something like strong complementarity, but now realized in a single quantum theory.

\section*{Acknowledgements}
We have benefited enormously from discussions with many people about firewalls.  In particular Juan Maldacena was of invaluable help in discussing many aspects of this work from its inception to its completion.  We are also grateful Igor Klebanov for technical assistance with the brane setup and Joe Polchinski for suggesting using strings and in general being a good sport.  We additionally thank Scott Aaronson, Nima Arkani-Hamed, Tom Banks, Alex Belin, Adam Brown, Raphael Bousso, Borun Chowdhury, Xi Dong, Liam Fitzpatrick, Ben Freivogel, Daniel Gottesman, Tom Hartman, Simeon Hellerman, Jared Kaplan, Don Marolf, Alex Maloney, Jonathan Maltz, Jonathan Oppenheim, Don Page, Kyriakos Papadodimas, Greg Prisament, Suvrat Raju, Grant Salton, Steve Shenker, Eva Silverstein, Douglas Stanford, Leonard Susskind, Brian Swingle, Erik Verlinde, Herman Verlinde, John Watrous, Edward Witten and all of the participants of the Stanford Firewall Alert meeting for providing a stimulating and chaotic environment for discussing the ideas in this paper and many others as well.  Finally we both especially thank John Preskill, to whom we dedicate this paper on his 60th birthday.  DH spent several months at the beginning of this project studying John's excellent lecture notes, without which he would have been able to contribute little.  DH is supported by the Princeton Center for Theoretical Science. PH is supported by the Canada Research Chairs program, the Perimeter Institute, CIFAR, NSERC and ONR through grant N000140811249. The Perimeter Institute is supported by Industry Canada and Ontario's Ministry of Economic Development \& Innovation.
\appendix
%
\section{The Absorption Probability for a Nambu-Goto String}\label{stringapp}
In this section we compute the low-energy absorption probability for transverse oscillations of a string stretching down the extremal black three brane geometry \eqref{extbb}.  This has been previously been computed by Maldacena and Callan \cite{Callan:1997kz}; our method is the same as in \cite{igor1} for a massless scalar.  The Nambu-Goto action is 
\be
S_{NG}=-\frac{1}{2\pi \ell_s^2}\int d^2\sigma \sqrt{-\det\left(G_{MN}\partial_iX^M\partial^i X^N\right)},
\ee
and parametrizing the string in static gauge and considering only oscillations in the $\mathbb{S}^5$ direction we have
\begin{align}\nonumber
t&=\tau\\\nonumber
r&=\sigma\\\nonumber
\vec{x}&=0\\\nonumber
\theta&=\theta(\tau,\sigma).
\end{align}
Linearizing the action in $\theta$ we find
\be
S=\frac{1}{4\pi\ell_s^2}\int d\sigma d\tau \sigma^2\left(Z(\sigma)\dot{\theta}^2-\theta'^2\right),
\ee
and looking at modes of definite frequency $\omega$ the equation of motion is
\be
\theta''+\frac{2}{\sigma}\theta'+Z(\sigma)\omega^2\theta=0.
\ee
Defining $\rho=\omega \sigma$, the equation becomes
\be\label{theq}
\theta''+\frac{2}{\rho}\theta'+\left(1+\left(\frac{\omega R}{\rho}\right)^4\right)\theta=0.
\ee
For $\rho \gg \omega R$ the solution is approximately
\be\label{bigthet}
\theta=A\frac{e^{i\rho}}{\rho}+B\frac{e^{-i\rho}}{\rho},
\ee
while for $\rho \ll \omega R$ the solution is approximately
\be\label{smallthet}
\theta=\tilde{A}e^{\frac{i\omega^2R^2}{\rho}}+\tilde{B}e^{\frac{-i\omega^2R^2}{\rho}}.
\ee
When $\omega R\ll1$ we can also find an approximate solution for $(\omega R)^{4/3}\ll \rho \ll 1$ by keeping only the derivative terms in \eqref{theq}:
\be
\theta=\frac{C}{\rho}+D.
\ee
Since this range overlaps with the other two ranges, we can use this solution to connect them together.  Since we are computing an absorption probability we want $\tilde{B}=0$, which means that matching the two ``inner'' regions gives 
\begin{align}\nonumber
C=i\omega^2 R^2\tilde{A}\\
D=\tilde{A}.
\end{align}
Matching the ``outer'' two regions gives
\begin{align}\nonumber
C=A+B\\
D=i(A-B),
\end{align}
so when $\omega R\ll1$ we have
\be
A\approx -B\approx-\frac{i}{2}\tilde{A}.
\ee
Finally to compute the absorption probability we inspect equations \eqref{bigthet} and \eqref{smallthet}, switching back from $\rho$ to $r$, and compute the square of the ratio of the coefficients of the waves.  The result is
\be
P_{abs}\sim (\omega R)^2,
\ee
consistent with \cite{Callan:1997kz}.  One could also study oscillations along the brane direction, according to \cite{Savvidy:1999wx} these give an absorption probability proportional to $(\omega R)^4$.
\bibliographystyle{jhep}
\bibliography{bibliography}
\end{document}